\pdfoutput=1

\documentclass[11pt,twoside,a4paper,cmspaper,final,collab]{cms-tdr}
\def\svnVersion{430352P}\def\cmsCernNoTag{CERN-EP-2016-292}\def\cmsCernDate{\today}\def\cmsMessage{Published in the Journal of High Energy Physics as \href{http://dx.doi.org/10.1007/JHEP10(2017)076}{\doi{10.1007/JHEP10(2017)076.}}}

\begin{document}\cmsNoteHeader{HIG-16-015}

\hyphenation{had-ron-i-za-tion}
\hyphenation{cal-or-i-me-ter}
\hyphenation{de-vices}
\RCS$Revision: 420101 $
\RCS$HeadURL: svn+ssh://svn.cern.ch/reps/tdr2/papers/HIG-16-015/trunk/HIG-16-015.tex $
\RCS$Id: HIG-16-015.tex 420101 2017-08-08 07:46:10Z ccaillol $

\newcommand{\mt}{\ensuremath{m_{\mathrm{T}}}\xspace}
\newcommand{\mH}{\ensuremath{125}\xspace}  %GHM
\newcommand{\lbound}{\ensuremath{25}\xspace}
\newcommand{\hbound}{\ensuremath{62.5}\xspace}
\newcommand{\mfour}{\ensuremath{m_{\mu\mu\PQb\PQb}}\xspace}
\newcommand{\Hb}{\ensuremath{{\mathrm{h}}}\xspace}
\newcommand{\ab}{\ensuremath{\mathrm{a}}\xspace}
\newcommand{\ma}{\ensuremath{m_{\ab}}\xspace}
\newcommand{\mb}{\ensuremath{m_{\PQb}}\xspace}
\newcommand{\mmumu}{\ensuremath{m_{\mu\mu}}\xspace}
\newcommand{\processmmbb}{\ensuremath{{\Hb}\to{\ab\ab}\to2\PGm2{\PQb}}\xspace}
\newcommand{\processmmtt}{\ensuremath{{\Hb}\to{\ab\ab}\to2\PGm2\PGt}\xspace}
\newcommand{\processtttt}{\ensuremath{{\Hb}\to{\ab\ab}\to4\PGt}\xspace}
\newcommand{\processmmmm}{\ensuremath{{\Hb}\to{\ab\ab}\to4\PGm}\xspace}
\newcommand{\Etmiss}{\ensuremath{\ET^\text{miss}}\xspace}
\newcommand{\Vptmiss}{\ensuremath{{\vec p}_{\mathrm{T}}^{\,\text{miss}}}\xspace}

\newlength\cmsFigWidth
\ifthenelse{\boolean{cms@external}}{\setlength\cmsFigWidth{0.4\columnwidth}}{\setlength\cmsFigWidth{0.4\textwidth}}
\ifthenelse{\boolean{cms@external}}{\providecommand{\cmsLeft}{top\xspace}}{\providecommand{\cmsLeft}{left\xspace}}
\ifthenelse{\boolean{cms@external}}{\providecommand{\cmsRight}{bottom\xspace}}{\providecommand{\cmsRight}{right\xspace}}
\cmsNoteHeader{HIG-16-015}
\title{Search for light bosons in decays of the 125\GeV Higgs boson in proton-proton collisions at $\sqrt{s}=8\TeV$}

\date{\today}
\abstract{
A search is presented for decays beyond the standard model of the 125\GeV Higgs bosons to a pair of light bosons, based on models with extended scalar sectors. Light boson masses between 5 and 62.5\GeV are probed in final states containing four $\PGt$ leptons, two muons and two \PQb quarks, or two muons and two $\PGt$ leptons. The results are from data in proton-proton collisions corresponding to an integrated luminosity of 19.7\fbinv, accumulated by the CMS experiment at the LHC at a center-of-mass energy of 8\TeV. No evidence for such exotic decays is found in the data. Upper limits are set on the product of the cross section and branching fraction for several signal processes. The results are also compared to predictions of two-Higgs-doublet models, including those with an additional scalar singlet.
}
\hypersetup{
pdfauthor={CMS Collaboration},%
pdftitle={Search for light bosons in decays of the 125 GeV Higgs boson in proton-proton collisions at sqrt(s) = 8 TeV},%
pdfsubject={CMS},%
pdfkeywords={CMS, physics, software, computing}}

\maketitle

\section{Introduction}
\label{sec:intro}
Studies of the recently discovered spin-0 particle $\Hb$~\cite{Aad:2012tfa,Chatrchyan:2012ufa,Chatrchyan:2013lba}, with a mass of 125\GeV and with properties consistent with the standard model (SM) Higgs boson~\cite{Khachatryan:2016vau}, severely constrain SM extensions that incorporate scalar sectors~\cite{Khachatryan:2014jba,Aad:2015gba,Aad:2015pla}.
There are many well-motivated models that predict the existence of decays of the Higgs boson to non-SM particles~\cite{PhysRevD.90.075004}.
Without making assumptions about the $\Hb(125)$ couplings to quarks, leptons, and vector bosons, other than that the scalar sector is composed only of doublets and singlets, the ATLAS and CMS collaborations at the CERN LHC exclude at a 95\% confidence level (CL) branching fractions of the Higgs boson to beyond SM (BSM) particles, $\mathcal{B}(\Hb\to\rm BSM)$, greater than 49\% and 52\%, respectively~\cite{Aad:2015gba,Khachatryan:2014jba}. Branching fractions as low as 34\% can be excluded at 95\% CL by combining the results obtained by the two experiments~\cite{Khachatryan:2016vau,Aad:2015zhl}.
The LHC experiments are expected to be able to constrain branching
fractions to new particles beyond the 5-10\% level using indirect measurements~\cite{Peskin:2012we,CMS:2013xfa,ATLAS:2013hta}.
In this context, it is interesting to explore the possibility of decays of the SM-like Higgs particle to lighter scalars or pseudoscalars~\cite{Dermisek:2005ar,Dermisek:2006wr,Chang:2008cw,PhysRevD.90.075004}.

The SM Higgs boson has an extremely narrow width
relative to its mass, because of its exceedingly small Yukawa couplings
to the SM fermions it can decay to. This suggests that any non-SM final state is likely to have a larger
partial width, and therefore a non-negligible branching fraction, compared to decays to SM particles~\cite{PhysRevD.90.075004}.
Examples of BSM models that provide such additional decay modes include those in which the Higgs boson serves as a portal to hidden-sector particles (e.g. dark matter)
that can couple to SM gauge bosons and fermions~\cite{portal}. Other models have extended scalar sectors, such as those proposed in two-Higgs-doublet models (2HDM)~\cite{2HDM-Lee,2HDM-Deshpande,2HDM-Haber,Gunion:1989we,Branco:2011iw}, in the next-to-minimal supersymmetric model (NMSSM)~\cite{Fayet:1974pd,Fayet:1977yc}, or in other models
in which a singlet Higgs field is added to the SM doublet sector.
The NMSSM is particularly well motivated as it provides a solution to the $\mu$ problem associated with supersymmetry breaking, and can provide a contribution to electroweak baryogenesis~\cite{Kim:1983dt,Ellwanger:2009dp}.

Both 2HDM and NMSSM may contain a light enough pseudoscalar state ($\ab$), which can yield a large $\Hb\to\ab\ab$ branching fraction.
In 2HDM, the mass of the pseudoscalar boson $\ab$ is a free parameter, but, if $\ma<m_{\Hb}/2$, fine-tuning of the 2HDM potential is required to keep the branching fraction $\mathcal{B}(\Hb\to\ab\ab)$ consistent with LHC data~\cite{Bernon:2014nxa}.
In NMSSM, there are two pseudoscalar Higgs bosons, $\ab_1$ and $\ab_2$. Constraints from the Peccei--Quinn~\cite{Peccei:1977hh,Peccei:1977ur} and $R$~\cite{Fayet:1976et,Fayet:1977yc} symmetries imply that the lighter $\ab_1$ is likely to have a mass smaller than that of the $\Hb$ boson~\cite{Ellwanger:2009dp}, and, since it is typically a singlet, suppression of $\mathcal{B}(\Hb\to \ab_1\ab_1)$ to a level compatible with observations is a natural possibility.
The minimal supersymmetric model (MSSM) contains a single pseudoscalar ($\mathrm{A}$), but the structure of the MSSM Higgs potential is such that its
mass cannot be below about 95\GeV when the scalar (to be identified with $\Hb$) has mass close to 125\GeV and is SM-like as implied by the LHC data~\cite{Heinemeyer:2011aa}. The phenomenology of decays of the observed SM-like Higgs boson to a pair of lighter Higgs bosons is detailed in Refs.~\cite{Celis:2013rcs,Grinstein:2013npa,Coleppa:2013dya,Chen:2013rba,Craig:2013hca,Wang:2013sha,PhysRevD.90.075004,Baglio:2014nea,Dumont:2014wha} for 2HDM, in Refs.~\cite{King:2012tr,Cao:2013gba,Christensen:2013dra,Cerdeno:2013cz,PhysRevD.90.075004} in the context of NMSSM or NMSSM-like, and in Refs.~\cite{Chalons:2012qe,Ahriche:2013vqa,PhysRevD.90.075004} in the general case of adding a singlet field to the SM or to a 2HDM prescription.

The 2HDM contains two Higgs doublet fields, $\Phi_1$ and $\Phi_2$, which, after symmetry breaking, lead to five physical states.
One of the free parameters in the 2HDM is $\tan\beta$, the ratio between the vacuum expectation values for the two doublets, expressed as $\tan\beta=v_2/v_1$.
The lightest scalar of the 2HDM is compatible with the
SM-like properties of the discovered boson in the limit where the other scalars all have large masses (decoupling limit),
and also in the alignment limit~\cite{Bernon:2015qea}, in which the neutral Higgs boson mass eigenstate is approximately aligned with the
direction of the vacuum expectation values for the scalar field. Approximate alignment, which is sufficient for consistency with LHC data, is possible for
a large portion of parameter space~\cite{Bernon:2015qea}, particularly when the pseudoscalar boson has sufficiently small mass to make $\Hb\to\ab\ab$ decays possible.

At lowest order, there are four types of 2HDM without flavor-changing neutral currents (FCNC), which can be characterized through the coupling of each fermion to the doublet structure, as shown in Table~\ref{tab:2hdm}.
The ratios of the Yukawa couplings of the pseudoscalar boson of the 2HDM relative to those of the
Higgs boson of the SM are functions of $\tan\beta$ and of the type of 2HDM, and are given in Table~\ref{tab:2hdm2}.
Type-1 and type-2 models are the ones commonly considered, and the latter are required in supersymmetric models. In these two cases, the leptons
have the same couplings as the down-type quarks. In type-3 2HDM, all quarks couple to $\Phi_2$ and all leptons couple to $\Phi_1$,
with the result that all leptonic or quark couplings of the pseudoscalar $\ab$ are proportional to $\tan\beta$ or $\cot\beta$, so that for large $\tan\beta$
the leptonic decays of $\ab$ dominate.

As implied previously, a complex $\textrm{SU}(2)_{\mathrm{L}}$ singlet field $S$ can be added to 2HDM; such models are called 2HDM+S, and include the NMSSM as a special case.
If $S$ mixes only weakly with the doublets, one of the CP-even scalars can again have SM-like properties. The addition of the singlet $S$ leads to two
additional singlet states, a second CP-odd scalar and a third CP-even scalar, which inherit a mixture of the fermion interactions of the Higgs doublets.
After mixing among the spin-0 states, the result is two CP-odd scalars, $\ab_{1}$ and $\ab_{2}$, and three CP-even scalars, $\Hb_{1}$, $\Hb_{2}$, and $\Hb_{3}$.
Of the latter, one can be identified with the observed SM-like state, $\Hb$. The branching fraction of the $\Hb$ boson to a pair of CP-even or CP-odd
bosons can be sizeable, leading to a wide variety of possible exotic $\Hb$ decays.
\begin{table}[!htbp]
\renewcommand{\arraystretch}{1.1}
\begin{center}
\topcaption{Doublets to which the different types of fermions couple in the four types of 2HDM without FCNC at lowest order.}\label{tab:2hdm}
\begin{tabular}{c|c|c|c|c}
& Type-1 & Type-2 & Type-3 (lepton-specific) & Type-4 (flipped) \\
\hline
Up-type quarks & $\Phi_2$ & $\Phi_2$ & $\Phi_2$ & $\Phi_2$ \\
Down-type quarks & $\Phi_2$ & $\Phi_1$ & $\Phi_2$ & $\Phi_1$ \\
Charged leptons & $\Phi_2$ & $\Phi_1$ & $\Phi_1$ & $\Phi_2$
\end{tabular}
\end{center}
\end{table}
\begin{table}[!htbp]
\renewcommand{\arraystretch}{1.1}
\begin{center}
\topcaption{Ratio of the Yukawa couplings of the pseudoscalar boson $\ab$ of the 2HDM relative to those of the Higgs boson of the SM, in the four types of 2HDM without FCNC at lowest order.}\label{tab:2hdm2}
\begin{tabular}{c|c|c|c|c}
& Type-1 & Type-2 & Type-3 (lepton-specific) & Type-4 (flipped) \\
\hline
Up-type quarks & $\phantom{-}\cot\beta$ & $\cot\beta$ & $\phantom{-}\cot\beta$ & $\phantom{-}\cot\beta$ \\
Down-type quarks & $-\cot\beta$ & $\tan\beta$ & $-\cot\beta$ & $\phantom{-}\tan\beta$ \\
Charged leptons & $-\cot\beta$ & $\tan\beta$ & $\phantom{-}\tan\beta$ & $-\cot\beta$
\end{tabular}
\end{center}
\end{table}
In the 2HDM and its extensions, the ratio of the decay widths of
a pseudoscalar boson to different types of leptons depends only on the masses of these leptons. In particular, for decays into muons and $\PGt$ leptons, and
a pseudoscalar boson of mass $\ma$, we can write~\cite{PhysRevD.90.075004,anatomy}:
\begin{equation}\label{eq:2hdm}
\frac{\Gamma(\ab\to\mu^+\mu^-)}{\Gamma(\ab\to\PGt^+\PGt^-)}=\frac{m_{\mu}^2\sqrt{1-(2m_{\mu}/\ma)^2}}{m_{\PGt}^2\sqrt{1-(2m_{\PGt}/\ma)^2}}.
\end{equation}
This kind of relation can also be written for electrons and muons.
In models where the pseudoscalar boson $\ab$ decays only to leptons,
its branching fraction to $\PGt$ leptons is greater than 99\% for pseudoscalar boson masses above 5\GeV.  This is a good approximation for pseudoscalar masses below twice the
bottom quark mass, or for type-3 2HDM, assuming loop-induced decays such as $\ab \to \Pg\Pg$ are ignored.  In type-1 and -2, and their extensions, a similar
relation exists between the partial decay widths of the pseudoscalar boson to leptons and to down-type quarks, for example, for muons and \PQb quarks, we can write~\cite{PhysRevD.90.075004,anatomy}:
\begin{equation}\label{eq:2hdm1}
\frac{\Gamma(\ab\to\mu^+\mu^-)}{\Gamma(\ab\to \PQb\PAQb)}=\frac{m_{\mu}^2\sqrt{1-(2m_{\mu}/\ma)^2}}{3m_{\PQb}^{2}\sqrt{1-(2m_{\PQb}/\ma)^2}\, (1+\textrm{QCD corrections})}.
\end{equation}

The factor of three in the denominator reflects the number of \PQb quark colors, and perturbative quantum chromodynamic (QCD) corrections are typically ${\simeq}20\%$~\cite{PhysRevD.90.075004}.
In models of type-3 or -4, however, the ratio of the partial decay widths depends on $\tan\beta$.

Three searches for decays of the 125\GeV Higgs boson to pairs of lighter scalars or pseudoscalars are described in this paper, where, for notational simplicity, the symbol $\ab$ refers to both the light scalar and light pseudoscalar:
\begin{itemize}
\item $\processtttt$,
\item $\processmmbb$,
\item $\processmmtt$.
\end{itemize}

The first analysis focuses on light boson masses above twice the $\PGt$ mass, using dedicated techniques to reconstruct the Lorentz-boosted $\PGt$ lepton pairs.
The two other analyses focus on masses large enough that the decay products are well separated from each other, and below
half of the Higgs boson mass. The production of the Higgs boson is assumed to be SM-like.
The results of these searches are interpreted in the 2HDM and 2HDM+S contexts, together with the two other analyses described in greater detail in the references given below:
\begin{itemize}
\item $\Hb\to\ab\ab\to4\PGm$~\cite{Khachatryan:2015wka};
\item $\processtttt$, using a different boosted $\PGt$ lepton reconstruction technique than the analysis with the same final state listed above~\cite{CMStttt1}.
\end{itemize}

These analyses are based on proton-proton collision data corresponding to an integrated luminosity of 19.7 fb$^{-1}$, recorded by the CMS experiment at the LHC at a center-of-mass energy of 8\TeV.
The D0 Collaboration at the Fermilab Tevatron published results for $\processmmtt$ and $\processmmmm$ searches for pseudoscalar masses $\ma$ between
3.5 and 19\GeV~\cite{Abazov:2009yi}, while ATLAS reported a search for $\processmmtt$ decays with $\ma$ between 3.7 and 50\GeV,
using special techniques to reconstruct Lorentz-boosted $\PGt$ lepton pairs~\cite{Aad:2015oqa}.
Additionally, CMS performed searches  for direct production of light pseudoscalars with mass between 5.5 and 14\GeV that decay to pairs of muons~\cite{Chatrchyan:2012am},
and with mass between 25 and 80\GeV that decay to pairs of $\PGt$ leptons~\cite{Khachatryan:2015baw}.

\section{The CMS detector, event simulation, and reconstruction}
\label{sec:cms-evt-reco}

The central feature of the CMS apparatus is a superconducting solenoid of 6\unit{m} internal diameter, providing an axial
magnetic field of 3.8\unit{T}. Within the solenoid volume are a silicon pixel and strip tracker,
a lead tungstate crystal electromagnetic calorimeter (ECAL), and a brass and scintillator hadron calorimeter (HCAL),
each composed of a barrel and two endcap sections. Extensive forward calorimetry complements the coverage provided by the barrel and endcap detectors. Muons are detected in gas-ionization chambers embedded in the steel
flux-return yoke outside the solenoid.

The first level of the CMS trigger system, composed of specialized hardware processors, uses information from the calorimeters and muon detectors to select the most interesting events in a fixed time interval of less than 4\mus. The high-level trigger processor farm further decreases the event rate from around 100\unit{kHz} to less than 1\unit{kHz}, before data storage.
A detailed description of the CMS detector, together with a definition of the coordinate system used and the relevant kinematic variables, can be found in Ref.~\cite{Chatrchyan:2008zzk}.

Samples of simulated events are used to model signal and background processes.
Drell-Yan, $\PW$+jets, \ttbar, and diboson events are simulated with \MADGRAPH 5.1.3.30~\cite{madgraph} using the matrix element calculation at leading-order (LO) precision in QCD; \PYTHIA 6.426~\cite{Sjostrand:2006za}
is used for parton showering, hadronization, and most particle decays; and \TAUOLA 27.121.5~\cite{Was200196}
is used specifically for $\PGt$ lepton decays. Single top quark events produced in association with a $\PW$ boson are generated using
\POWHEG 1.0 r1380~\cite{Alioli:2009je,Alioli:2010xd,Re:2010bp,Frixione:2007vw}, interfaced to \PYTHIA for parton showering.
Signal samples are generated with \PYTHIA using its built-in 2HDM and NMSSM generator routines.
Background and signal samples use the {\sc CTEQ6L}~\cite{Pumplin:2002vw} parton distribution functions (PDFs).
Minimum-bias collision events generated with \PYTHIA are added to all Monte Carlo (MC) samples to reproduce the observed concurrent pp collisions in each bunch crossing (pileup).
The average number of pileup interactions in 2012 data was 20.
All generated events are passed through the full {\sc Geant4}~\cite{Agostinelli:2002hh,1610988}
based simulation of the CMS apparatus and are reconstructed
with the same CMS software that is used to reconstruct the data.

Event reconstruction relies on a particle-flow (PF) algorithm, which combines information from different subdetectors to reconstruct
individual particles~\cite{CMS-PAS-PFT-09-001,PFT-10-001}: neutral and charged hadrons, photons, electrons, and muons. More complex objects are reconstructed by combining the PF candidates. A deterministic annealing algorithm~\cite{DAA1,DAA2} is used to reconstruct the collision vertices. The vertex with the maximum sum in the squared transverse momenta ($\PT^2$) of all associated charged particles is defined as the primary vertex. The longitudinal and radial distances of the vertex from the center of the detector
must be smaller than 24 and 2\unit{cm}, respectively.

Muons are reconstructed by matching hits in the silicon tracker and in the muon system~\cite{MuonReco}.
Global muon tracks are fitted from hits in both detectors. A preselection is applied to the global muon tracks, with requirements
on their impact parameters, to suppress non-prompt muons produced from the $\Pp\Pp$~collision or muons from cosmic rays.

Electrons are reconstructed from groups of one or more associated clusters of energy deposited in the ECAL.
Electrons are identified through a multivariate (MVA) method~\cite{TMVA} trained to discriminate electrons from quark and gluon jets~\cite{EleMVA}.

The muon and electron relative isolation is defined as:
\begin{equation}
I_{\mathrm{rel}}=\left[ \sum\limits_{\mathrm{charged}} \pt + \mathrm{max} \left(0, \sum\limits_{\mathrm{neutral}} \pt + \sum\limits_{\gamma} \pt -\frac{1}{2} \sum\limits_{\mathrm{charged,PU}} \pt\right)\right] / \pt,
\label{eq:Iso}
\end{equation}
where $\Sigma_{\mathrm{charged}} \pt$ is the sum of the magnitudes of the transverse momenta of charged hadrons, electrons and muons originating from the primary vertex,
$\Sigma_{\mathrm{neutral}} \pt$ is the corresponding sum
for neutral hadrons and $\Sigma_{\gamma}$ for photons, and  $\Sigma_{\mathrm{charged,PU}} \pt$  is the sum of the transverse momentum
of charged hadrons, electrons, and muons originating from other reconstructed vertices. The particles considered in the isolation calculation
are inside a cone with a radius
$\Delta R = \sqrt{\smash[b]{(\Delta\eta)^2 + (\Delta\phi)^2}}$ = 0.4 around the lepton direction, where $\Delta\eta$ and $\Delta\phi$ are the differences of
pseudorapidity and azimuthal angle in radians between the particles and the lepton direction, respectively. The factor $\frac{1}{2}$ originates from the approximate ratio of the neutral to charged candidates in a jet. In the search for $\processtttt$, the isolation criteria are extended to veto the presence of reconstructed leptons within the $\Delta R$ = 0.4 cone, as detailed in Section~\ref{sec:tttt}.

Jets are reconstructed by clustering charged and neutral particles using an anti-\kt algorithm~\cite{antikt}, implemented in the \FASTJET library~\cite{Cacciari:2011ma,Cacciari:fastjet2}, with a distance parameter of 0.5.
The reconstructed jet energy is corrected for effects from the detector response as a function of the jet $\pt$ and $\eta$.
Furthermore, contamination from pileup, underlying events, and electronic noise is subtracted on a statistical basis~\cite{CMSJetPaper}.
An eta-dependent tuning of the jet energy resolution in the simulation is performed to match the resolution observed in data~\cite{CMSJetPaper}. The combined secondary vertex (CSV) algorithm is used to identify jets that are likely to
originate from a $\PQb$ quark ("\PQb jets"). The algorithm exploits the track-based lifetime information together with the secondary vertices associated with
the jet to provide a likelihood ratio discriminator for the $\PQb$ jet identification~\cite{1748-0221-8-04-P04013}. A set of $\pt$-dependent correction factors are applied to simulated events to account for differences in the $\PQb$ tagging efficiency between data and simulation~\cite{1748-0221-8-04-P04013}.

Tau leptons that decay into a jet of hadrons and a neutrino, denoted $\PGt_{\text{h}}$, are identified with a hadron-plus-strips (HPS) algorithm, which matches tracks and ECAL energy deposits
to reconstruct $\PGt$ candidates in one of the one-prong, one-prong + $\PGpz$(s), and three-prong decay modes~\cite{TauID}.
Reconstructed $\PGt_{\text{h}}$ candidates are seeded from anti-\kt jets with a distance parameter of 0.5.  For each jet, $\PGt$ candidates are constructed
from the jet constituents according to criteria that include consistency with the vertex of the hard interaction and
consistency with the $\PGpz$ mass hypothesis. Two methods for rejecting quark and gluon jets are employed, depending on the analysis.  The first is a straightforward selection based on the isolation variable, while the second uses a multivariate analysis (MVA) discriminator
that takes into account variables related to the isolation, to the transverse impact parameter of the leading track of the $\PGt_{\text{h}}$ candidate, and to the distance between the $\PGt$ production point and the decay vertex in the case of three-prong decay modes~\cite{TauID}.
MVA-based discriminators are implemented to reduce the rates at which electrons or muons are misidentified as $\Pgt_{\text{h}}$ candidates. Muons or electrons from leptonic decays of $\PGt$ leptons are indistinguishable from prompt leptonic decay products of $\PW$ and $\PZ$ bosons and are reconstructed as described earlier.

The missing transverse energy, $\Etmiss$, is defined as the magnitude of \Vptmiss,
which is the negative sum of $\vec{p}_{\rm T}$ of all PF candidates.
The jet energy calibration introduces corrections to the $\Etmiss$ measurement. The $\Etmiss$ significance variable, which estimates the compatibility of the reconstructed $\Etmiss$ with zero, is calculated via a likelihood function on an event-by-event basis~\cite{1748-0221-6-09-P09001}.

As part of the quality requirements, events in which an abnormally high level of noise is detected in the HCAL barrel or endcap detectors are rejected~\cite{Apresyan:2015kla}.

\section{Search for \texorpdfstring{$\processtttt$ decays}{Higgs decay to two di-tau pairs}}
\label{sec:tttt}

This analysis considers 4$\PGt$ final states arising from $\processtttt$ decay, where the Higgs boson is produced via gluon fusion ($ \cPg\cPg\Hb $), in association with a $\PW$ or \Z boson ($ \PW\Hb $ or $ \PZ\Hb $), or via vector boson fusion (VBF). Light boson masses are probed in the range 5\mdash15\GeV, where the branching fraction of the light boson to $\PGt$ leptons is expected to be large in certain 2HDM models.  To illustrate the performance of the analysis, a mass of 9\GeV is chosen as a benchmark model throughout this section; it represents a type-2 2HDM variant in which the pseudoscalar branching fraction to $\PGt$ leptons is dominant.  The large Lorentz boost of the $\ab$ boson at such light masses causes its decay products to overlap.  To maximize the sensitivity to overlapping $\PGt$ leptons, a special boosted $\PGt\PGt$ pair reconstruction technique is employed, based on the specific final state in which one $\PGt$ lepton decays to a muon.  This analysis is performed in two search regions based on the transverse mass ($\mt$) formed from a high-\pt muon and the $\ptmiss$. These two regions are designed to distinguish between the $ \PW\Hb $ production mode and other modes (primarily $ \cPg\cPg\Hb $) without significant $\Etmiss$.

\subsection{Event selection}

Events considered in this search are selected with a single muon trigger that requires the presence of an isolated muon with $ \pt > 24\GeV $ and $|\eta| < 2.1$.
This analysis specifically targets the event topology with one isolated high \pt muon, and at least one boosted $\Pgt\Pgt$ pair in which one $\Pgt$ lepton decays to a muon and neutrinos ($\Pgt_\Pgm$).
No assumption is made on the decay of the second $\Pgt$ lepton in the boosted $\Pgt\Pgt$ pair.
Because of the features of this topology, it is convenient to define the ``trigger muon'' candidate, $\Pgm_{\text{trg}}$, referring to the isolated high $\pt$ muon triggering the event, and the ``$\Pgt_\Pgm\Pgt_\text{X}$ object'', aiming to reconstruct the decay products of the boosted $\Pgt\Pgt$ pair. This topology is characteristic of two classes of signal events:

\begin{enumerate}
\item The Higgs boson is produced through gluon fusion or vector boson fusion and decays as $\Hb\to\ab(\to\Pgt_\Pgm\Pgt_\text{X})\ab(\to\Pgt_\Pgm\Pgt_\text{X})$.  When the $\Pgt_\Pgm$ from the decay of one $\ab$ has both a high \pt and is well separated from the $\Pgt_\text{X}$ arising from the same decay, it will satisfy the trigger muon criteria.  The other $\Pgt\Pgt$ pair is reconstructed as a $\Pgt_\Pgm\Pgt_\text{X}$ object.
\item The Higgs boson is produced through associated production with a \PW\xspace or a \Z boson that then decays to isolated muons. The Higgs boson decay considered here is $\Hb\to\ab(\to\Pgt_\Pgm\Pgt_\text{X})\ab(\to\Pgt_\text{X}\Pgt_\text{X})$.  The muon from the \PW\xspace or \Z decay is required to pass the trigger criteria, one of the $\Pgt\Pgt$ pairs is reconstructed as a $\Pgt_\Pgm\Pgt_\text{X}$ object, and no requirement is applied to the second $\Pgt\Pgt$ pair.
\end{enumerate}

The remainder of this subsection describes selection and reconstruction criteria for the muon that fires the trigger, and for the $\Pgt_\Pgm\Pgt_\text{X}$ object.

The reconstructed $\Pgm_{\text{trg}}$ object must be located within $\Delta R < 0.1$ of the isolated muon reconstructed in the trigger system.  It is also required to have $ \pt > 25\GeV $,
$|\eta| <$ 2.1, be well reconstructed in both the muon detectors and the silicon tracker, have a high-quality track fit, and be consistent with originating
from the primary $\Pp\Pp$ interaction in the event.  In addition, it must be isolated from other photons, hadrons, and leptons in the detector.
Isolation from photons and hadrons is enforced by requiring that the muon relative isolation, as defined in Eq.~(\ref{eq:Iso}), is less than 0.12.  To be isolated from other leptons, the trigger muon is required to have no identified electrons (\pt $>$ 7\GeV, $|\eta| < 2.5$),
muons ($\PT > 5\GeV$, $|\eta| < 2.4$, passing $\PGt_{\PGm}$ criteria below),
or $\PGt$ leptons (\PT $>$ 10\GeV, $|\eta| < 2.3$, passing modified HPS criteria, as described below)
reconstructed within $\Delta R$ = 0.4 of the trigger muon direction.  The requirement of isolation from nearby leptons, in addition to the isolation requirement of Eq.~(\ref{eq:Iso}), ensures that a trigger muon originating from a $\PGt$ lepton decay, where the $\PGt$ lepton originates from a pseudoscalar decay, is well isolated from the other $\PGt$ lepton in the pseudoscalar decay pair.  In this way, the high level trigger and ``trigger muon'' identification criteria are efficient for low-\pt $\PGt$ decay muons expected to pass the trigger in the $ \cPg\cPg\Hb $ and VBF production modes, provided that $\PGt$ leptons from the pseudoscalar decay are well separated or one of the $\PGt$ leptons has \pt low enough not to affect the isolation of the other $\PGt$ lepton.  The isolation requirements are also efficient for high-\pt isolated muons from $\PW$ boson decays expected in the $ \PW\Hb $ associated production mode.

The muon from the $\PGt$ lepton decaying via the muon channel ($\PGt_{\PGm}$) is required to have \pt $>$ 5\GeV and $|\eta| <$ 2.4, be well reconstructed in the silicon tracker,
have a high-quality track fit, be consistent with originating from the primary vertex in the event, and be separated by at least
$\Delta R$ = 0.5 from the trigger muon.  Because no isolation requirement is placed on the $\PGt_{\PGm}$ candidate, it
can be identified with high efficiency in the presence of a nearby $\PGt$ lepton.  Overall, the trigger and $\PGt_{\PGm}$ quality criteria are similar,
but the $\PGt_{\PGm}$ criteria are optimized for low-\pt non-isolated muons, while the trigger muon criteria are optimized for
high-\pt isolated muons.

Since the final state in this analysis includes a pair of boosted $\PGt$ leptons from pseudoscalar decay, the HPS algorithm is
modified to maintain high efficiency for overlapping $\PGt$ leptons.
All jet constituents are checked for the presence of $\PGt_{\PGm}$ candidates as defined above.
Only jets that have at least one muon candidate passing the $\PGt_{\PGm}$ criteria among their constituents are used to seed the HPS reconstruction.
Within these selected jets, the muon is excluded from the set of jet constituents before running the HPS reconstruction algorithm. The HPS reconstruction then proceeds as
described in Section~\ref{sec:cms-evt-reco}, and the resulting $\PGt$ lepton is required to have \pt $>$ 20\GeV and $|\eta| <$ 2.3.
The combination of the $\PGt_{\PGm}$ and isolated HPS $\PGt$ candidates resulting from this selection form the $\PGt_{\PGm}\PGt_\text{X}$ object as it is designed to reconstruct boosted
$a\to\PGt_{\PGm}\PGt_\text{X}$ decays.  The HPS $\PGt$ candidate is referred to as $\PGt_\text{X}$ because no anti-electron or
anti-muon discriminators are applied to it; although $\PGt$ leptons decaying to electrons and muons can thus pass the HPS selection, the vast majority (${\simeq}97\%$)
of selected $\PGt$ candidates in simulated \Hb$\to\ab\ab$ samples are hadronically decaying $\PGt$ leptons.  The modified HPS $\PGt$ lepton reconstruction and isolation requirements have a similar efficiency for $\Hb\to \ab\ab$ decays as the standard HPS and isolation requirements have for $\PZ\to\PGt\PGt$ decays.

This analysis requires at least one $\PGt_{\PGm}\PGt_\text{X}$ object, which reconstructs a single $\ab\to\PGt\PGt$ decay, per event.  The $\PGt_{\PGm}\PGt_\text{X}$ object consists of a muon, one or three other charged particle tracks, and zero or more neutral hadrons,
and could therefore arise from misidentifying the decay products of a bottom quark jet.
To further distinguish $\PGt_{\PGm}\PGt_\text{X}$ objects from background, the seed jet of the HPS reconstructed $\PGt_\text{X}$ (excluding any identified $\PGt_{\PGm}$ candidate) is required not to be identified as a $\PQb$ jet.

\subsection{Signal and background estimation}

The main background contributions to this search arise from Drell-Yan dimuon pairs produced in association with jets,
($\PW\to\PGm\cPgn$) + jets, \ttbar with muons in the final state, and QCD multijet events.
In order to reduce the Drell-Yan background, the trigger muon and $\PGt_\text{X}$ candidates are required to have the same sign (SS) of electric charge.
To minimize backgrounds with jets misidentified as $\PGt$ candidates, the $\PGt_{\PGm}$ and $\PGt_{\text{X}}$ objects are required to have opposite sign.
The signal region is defined by events passing all the requirements described above, as well as $m_{\PGm+\text{X}}$ $\geq$ 4\GeV, where $m_{\PGm+\text{X}}$ is the invariant mass calculated from the four-vectors of the two components of the $\PGt_{\PGm}\PGt_\text{X}$ object.  The choice of 4\GeV reduces the expected background contribution by about 95\%, while keeping approximately 75\% of the expected events in the case of the $ \cPg\cPg\Hb $ benchmark 9\GeV pseudoscalar mass sample.
Signal acceptance is calculated from the simulated samples for masses between 5 and 15\GeV.
The expected signal acceptance is corrected using \pt- and $|\eta|$-dependent scale
factors to account for known differences in the $\PQb$ veto efficiency between data and simulation~\cite{1748-0221-8-04-P04013}.

Events are classified into two analysis bins depending on the value of the transverse mass between the trigger muon momentum
and the \Vptmiss, defined as
\begin{equation}
\mt = \sqrt{2\pt^{\PGm_{\text{trg}}}\Etmiss[1 - \cos{\Delta\phi(\PGm_{\text{trg}}, \Vptmiss)}]},
\label{eq:MT}
\end{equation}
where $\Delta\phi(\PGm_{\text{trg}}, \Vptmiss)$ is the azimuthal angle between the trigger muon position vector and \Vptmiss vector.
The contribution of signal events for the different production modes in the low-$\mt$ and high-$\mt$ bins for a representative pseudoscalar mass of 9\GeV, and
assuming $\mathcal{B}(\Hb\to \ab\ab) \, \mathcal{B}^{2}(\ab \to \PGt^{+}\PGt^{-})=0.1$, is given in Table~\ref{tab:yields_tttt}.  For $\mt \le$ 50\GeV, $ \cPg\cPg\Hb $ fusion production accounts for about 85\% of the expected signal, VBF accounts for another 10\%, and associated production accounts for the rest.  For $\mt >$ 50\GeV, $ \cPg\cPg\Hb $ and $ \PW\Hb $ productions each account for about 40\% of the expected signal and $ \PZ\Hb $ and VBF productions account for the rest. Dividing selected events in two $\mt$ categories increases the sensitivity to models (for example Ref.~\cite{Ferreira:2014naa}) where the $ \cPg\cPg\Hb $ production rate would be modified with respect to the SM expectation because of different Yukawa couplings of the fermions appearing in the loop, whereas the $ \PW\Hb $ and $ \PZ\Hb $ production rates would be similar as in the SM in the case of the alignment limit of 2HDM.
\begin{table}
\renewcommand{\arraystretch}{1.2}
\begin{center}
\topcaption{Expected signal yields for the \processtttt process for a representative pseudoscalar mass of 9\GeV, in both $\mt$ bins, assuming SM cross sections and $\mathcal{B}(\Hb\to \ab\ab) \, \mathcal{B}^{2}(\ab \to \PGt^{+}\PGt^{-})=0.1$, in the context of the $\processtttt$ search. Expected background yields as well as observed numbers of events are also quoted.
Only the statistical uncertainty is given for signal yields.}\label{tab:yields_tttt}
\begin{tabular}{c|c|c}
\multicolumn{1}{c|}{} & $\mt\leq 50\GeV$ & $\mt>50\GeV$ \\
\hline
$ \cPg\cPg\Hb $ & $4.6\pm0.3$ & $0.8\pm0.1$ \\
$ \PW\Hb $ & $0.27\pm0.02$ & $0.70\pm0.03$ \\
$ \PZ\Hb $ & $0.068\pm0.005$ & $0.19\pm0.01$ \\
VBF & $0.51\pm0.03$ & $0.09\pm0.01$ \\
\hline
SM background & $5.4\pm 1.0\stat ^{+4.2}_{-4.6}\syst $ & $6.1\pm 1.6\stat ^{+3.7}_{-3.6}\syst $ \\
\hline
Observed & 7 & 14
\end{tabular}
\end{center}
\end{table}

There are several mechanisms that result in $\PGt_{\PGm}\PGt_{\text{X}}$ misidentification, for example jets with semileptonic decays, jets with double semileptonic decays, or resonances in $\PQb$ or light-flavor jet fragmentation. It is impractical to simulate all backgrounds to the required statistical precision.
Therefore, the number of background events in the low-$\mt$ (high-$\mt$) signal region, denoted
$N^{\text{low-\mt(high-\mt)}}_{\text{bkg}}(m_{\PGm+\text{X}} \geq 4\GeV)$, is estimated independently from three event samples.  In each background estimation sample, the isolation energy around the $\PGt_{\text{X}}$ candidate is required to be between 1 and 5\GeV, as opposed to the signal sample requirement of isolation energy less than 1\GeV.  The three samples are:
\begin{enumerate}
\item Observed events passing all other signal selections;
\item Simulated Drell-Yan, $\PW$+jets, \ttbar, and diboson events passing all other signal selections;
\item Observed events passing all other signal selections, but with inverted $\PGm_{\text{trg}}$ relative isolation.
\end{enumerate}
The background estimate from each sample is normalized to match the observed data yield in the signal-free region with $m_{\PGm+\text{X}} < 2\GeV$.
The final background prediction in the low-$\mt$ (high-$\mt$) bin is taken as the arithmetic
mean of the estimates from the three background estimation samples with $\mt \leq 50\GeV(\mt > 50\GeV)$.
The positive (negative) systematic uncertainty is taken as the difference between the largest (smallest) of the
three plus (minus) its statistical uncertainty and the average.
In the low-$\mt$ bin, the background yield is estimated to be $5.4\pm 1.0\stat^{+4.2}_{-4.6}\syst$ events, while in the high-$\mt$ bin it is estimated to be $6.1\pm1.6\stat ^{+3.7}_{-3.6}\syst $ events. The uncertainty on the background yield is dominated by the limited statistical precision in the control samples, owing to the rare final state being probed.  This uncertainty is the dominant source of systematic error in the interpretation of the results of this search in terms of an upper limit on the branching fraction of the Higgs boson to light pseudoscalar states.

The relaxed $\PGt_{\text{X}}$ isolation requirement common to each sample implies that these background estimation samples should be enriched in events with jets.  Simulated samples of $\PW$+jets and \ttbar events, in which the $\PGt_{\PGm}\PGt_{\text{X}}$ candidate arises from misidentified jets, have been used to check that events with nonisolated $\PGt_{\text{X}}$ candidates have the same kinematic properties as those of the signal sample.

\begin{figure}[hbtp]
\begin{center}
\includegraphics[width=1.24\cmsFigWidth]{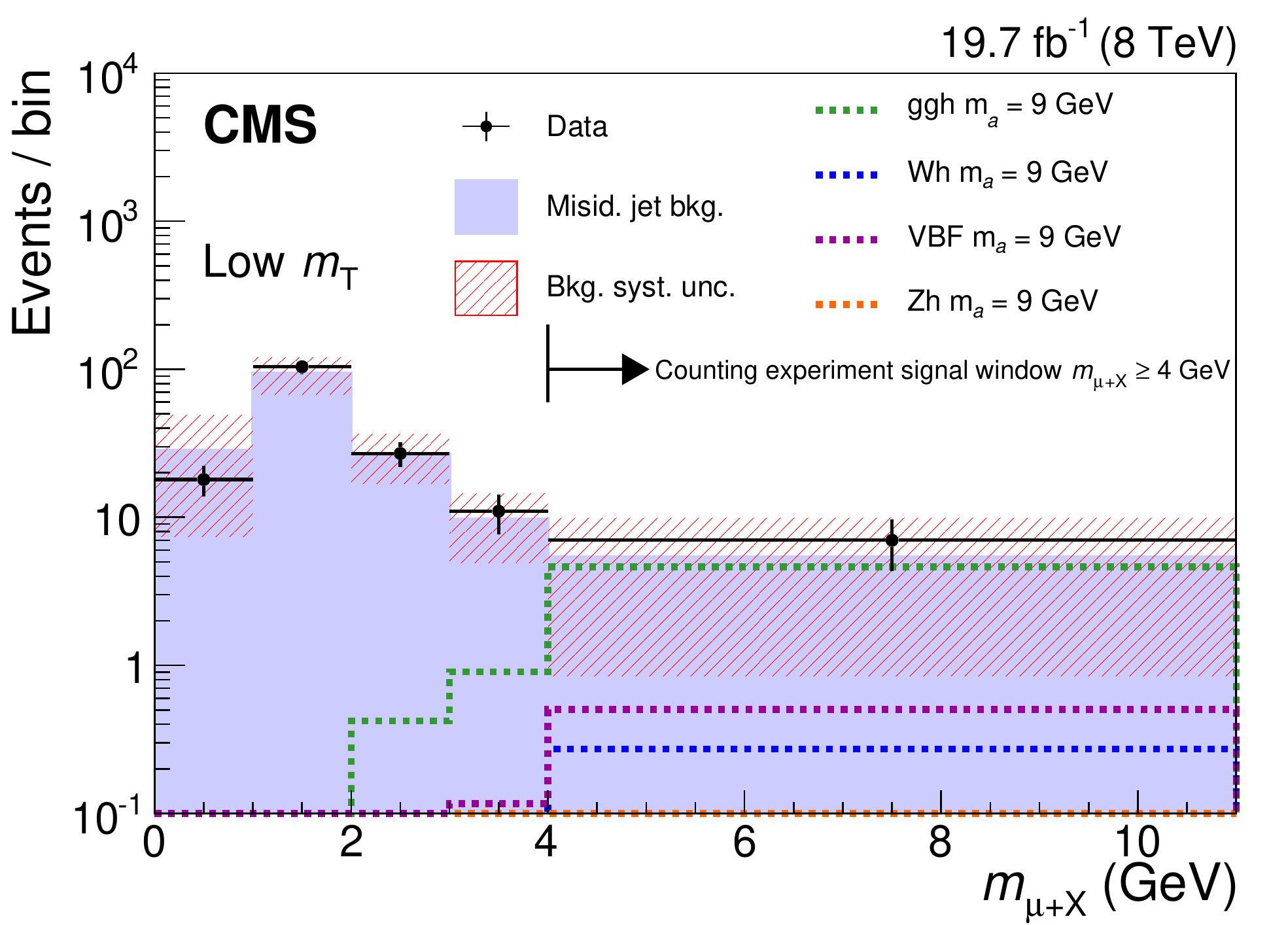}
\includegraphics[width=1.24\cmsFigWidth]{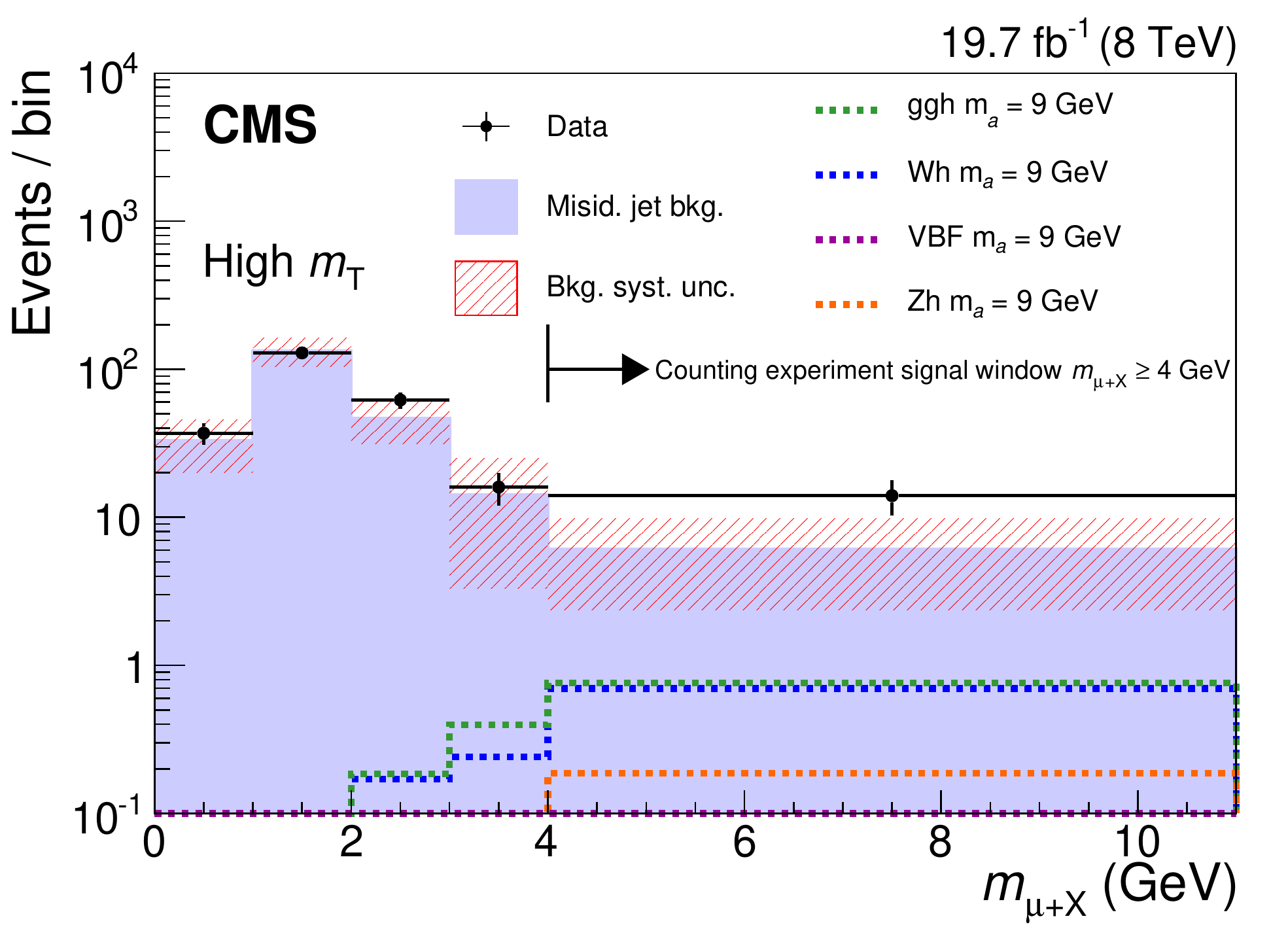}
\caption{Comparison, for the $\processtttt$ search, of $m_{\PGm+\text{X}}$ distributions for data (black markers) and the misidentified jet background estimate (solid histogram) in the low-$\mt$ (left) and high-$\mt$ (right) bins.  Predicted signal distributions (dotted lines) for each of the four Higgs boson production mechanisms are also shown; the distributions are normalized to an integrated luminosity of the data sample of 19.7\fbinv, assuming SM Higgs boson production cross sections and $\mathcal{B}(\Hb \to \ab\ab)\,\mathcal{B}^{2}(\ab \to \PGt^+\PGt^-) = 0.1$. The last bin on the right contains all the events with $m_{\PGm+\text{X}}\geq 4\GeV$, which correspond to the numbers
reported in Table \ref{tab:yields_tttt}.}
\label{fig:final_tttt}
\end{center}
\end{figure}

Figure~\ref{fig:final_tttt} shows the estimated misidentified jet background, the search region data, and simulations of the four signal production models for both $\mt$ bins.  Seven and fourteen events are observed in the low- and
high-$\mt$ bins, respectively.

\section{Search for \texorpdfstring{$\Hb \to \ab\ab\to2\PGm2\PQb$}{mmbb} decays}
\label{sec:mmbb}
The search for a new scalar in $\processmmbb$ decays is restricted to masses between 25 and 62.5\GeV. The upper bound is imposed by the kinematic constraint of $m_{\Ph}=125\GeV$, while there is a sensitivity loss for this search below the lower bound due to overlap between the two $\PQb$ jets or the two muons arising from an increased boost of the pseudoscalars~\cite{Curtin:2014pda}. A slightly wider pseudoscalar mass range is however used for the selection, the optimization aiming at maximum expected signal significance, and the eventual background modeling. In particular, the wider mass range ensures a good description of the background distribution over the entire search region, including regions near the boundaries. Events with an invariant mass $\mmumu$ outside the range 20-70\GeV are discarded.

\subsection{Event selection}
In the search for $\processmmbb$ decays, events are triggered based on the presence of two muons with $p_{\rm T}>17\GeV$ and $\pt>8\GeV$.
For the offline selection, the leading muon $\pt$ threshold is increased to 24\GeV, while the subleading muon $\pt$ must exceed 9\GeV. The two muon candidates are required to have opposite electric charges and to be isolated.
If more than one muon is found for a given sign, the one with the highest $\pt$ is selected.
At least two jets with $\pt>15\GeV$ and $|\eta|<2.4$ are required to satisfy \PQb-tag requirements that allow only $\mathcal{O}(1\%)$ of the light quark jets to survive, for an efficiency of ${\simeq}65\%$ for genuine \PQb jets. The $\Etmiss$ significance of the event has to be less than 6. Events outside the $|\mfour-\mH\GeV|<25\GeV$ window are discarded.

\subsection{Signal and background estimation}

As presented in Table~\ref{tab:yieldsmmbb}, the expected background yield estimated from simulation over the whole mass range considered is $235\pm35$ events, dominated by Drell-Yan events in the dilepton final state, followed by $\ttbar$ in dilepton decays, $\ttbar\,(\ell\ell)$. This should be compared with 252 events observed in data. To evaluate the signal yield, only the gluon fusion Higgs boson production mechanism with the next-to-leading-order (NLO) cross section of $\sigma_{\rm{gg}\Hb}\simeq 19.3\unit{pb}$~\cite{Dittmaier:2011ti} is considered. Other SM Higgs production modes are found to contribute less than 5\% to the signal yield and are neglected. Assuming a branching fraction of 10\% for $\Hb\to\ab\ab$ together with $\tanb=2$ in the context of type-3 2HDM+S, one can obtain $2  \mathcal{B}(\ab \to {\PQb}\overline{\PQb})\mathcal{B}(\ab \to \mu^+\mu^- )=1.7\times10^{-3}$ for $\ma=30\GeV$, where no strong dependence on $\ma$ is expected for $\mathcal{B}(\ab\to \rm{f}\overline{\rm{f}})$, with $\rm{f}$ being a muon or a \PQb quark~\cite{PhysRevD.90.075004}. In this scenario, about one signal event is expected to survive the event selection discussed earlier.
\begin{table}[!ht]
\renewcommand{\arraystretch}{1.1}
\begin{center}
\topcaption{Expected signal and background yields, together with the number of observed events, for the $\processmmbb$ search, in the range $20\leq\mmumu\leq 70\GeV$.
Signal yields are evaluated assuming $\mathcal{B}(\Hb\to\ab\ab)=10\%$ and $\mathcal{B}(\ab\ab\to\PGm^+\PGm^-\PQb\overline{\PQb})=1.7\times 10^{-3}$, with the latter obtained in the context of type-3 2HDM+S with $\tan\beta=2$. }
\begin{tabular}{l|c|c|c|c}
&\multicolumn{2}{c|}{${\PZ}/\gamma^*$+jets ($m_{\ell\ell} > 10\GeV$)}&$\ttbar\,(\ell\ell)$& Other     \\
\hline
Backgrounds                     &\multicolumn{2}{c|}{                  $210\pm35$}             &       $22\pm1$                &$3\pm1$                \\
\hline
Total                           &\multicolumn{4}{c}{ $235\pm35$}\\
Data                            &\multicolumn{4}{c}{ $252$}\\[-1.0ex]
\multicolumn{5}{c}{}\\
&       $\ma = 30\GeV$     &       $\ma = 40\GeV$                             &$\ma = 50\GeV$& $\ma = 60\GeV$\\
\hline
Signal                          &       1.18                            &       0.97                                                   &1.11                           &       1.49
\end{tabular}
\label{tab:yieldsmmbb}
\end{center}
\end{table}

The signal yield is extracted using a fit to the reconstructed $\mmumu$ distribution in data. The signal shape is modeled with a weighted sum of Voigt profile~\cite{NIST:DLMF} and Crystal Ball~\cite{SLAC-R-236} functions with a common mass parameter $\ma$,
\begin{linenomath}
\begin{equation}
\label{eq:signal}
{\rm S}(\mmumu|w,\sigma, \gamma,n,\sigma_{cb}, \alpha,\ma) \equiv w\, {\rm V}(\mmumu|\sigma, \gamma,\ma)+(1-w)\, {\rm CB}(\mmumu|n,\sigma_{ cb}, \alpha,\ma).
\end{equation}
\end{linenomath}
The Voigt profile function, ${\rm V}(\mmumu|\sigma, \gamma,\ma)$, is a convolution of Lorentz and Gaussian profiles
with $\gamma$ and $\sigma$ being the widths of the respective functions, both centered at $\ma$.
The Crystal Ball function, ${\rm CB}(\mmumu|n,\sigma_{cb}, \alpha,\ma)$, has a Gaussian core centered at $\ma$ with a width of $\sigma_{cb}$ together with a power-law low-end tail $A \, (B - ({\mmumu-\ma})/{\sigma_{cb}} )^{-n}$ below a certain threshold $\alpha$. The combination introduced in Eq.~(\ref{eq:signal}) is found to describe well the $\mmumu$ distribution in the simulated signal samples.

The initial values for the signal model parameters are extracted from a simultaneous fit of the model to simulated signal samples with different pseudoscalar masses. All parameters in the signal model are found to be independent of $\ma$ except $\sigma$ and $\sigma_{cb}$, which show a linear dependence.
The only floating parameter in these linear models are the slopes, $s_\sigma$ and $s_{\sigma_{cb}}$ for $\sigma$ and $\sigma_{cb}$, respectively.
The signal model with the three free parameters, $\ma$, $s_\sigma$ and $s_{\sigma_{cb}}$, is interpolated for mass hypotheses not covered by the simulated samples. The validity of the interpolation is checked within the $\left[\lbound,\hbound\right]\GeV$ range of the dimuon mass, and towards the boundaries.

The background is evaluated through a fit to the $\mmumu$ distribution in data. The shape for the background is modeled with a set of analytical functions, using the discrete profiling method~\cite{Dauncey:2014xga,Khachatryan:2014ira,Aad:2015zhl}. In this approach the choice of the functional form of the background shape is considered as a discrete nuisance parameter. This means that the likelihood function for the signal-plus-background fit has the form of
\begin{linenomath}
\begin{equation}
\label{eq:likelihood}
\mathcal{L}({\rm data}|\mu, \theta_\mu, b_\mu),
\end{equation}
\end{linenomath}
where $\mu$ is the measured quantity of signal, $\theta_\mu$ are the corresponding nuisance parameters, and $b_\mu$ are the different background functions considered. Therefore, the uncertainty associated with the choice of the background model is treated in a similar way as other uncertainties associated with continuous nuisance parameters in the fit. The space of the background model contains multiple candidate models:
different parametrizations of polynomials together with $1/P_n(x)$ functions where $P_n(x)\equiv x + \sum_{i=2}^{n}\,\alpha_ix^i$. The degree of polynomials in each category is determined through statistical tests to ensure the sufficiency of the number of parameters and to avoid overfitting the data~\cite{Khachatryan:2014ira}.
Starting from the lowest degree for every candidate model, the necessity to increase the degree of the polynomial is examined. The model candidate with the higher degree is fit to data and a $p$-value is evaluated according to the number of degrees of freedom and the relative uncertainty of the parameters. Candidates with $p$-values below 5\% are discarded.

The input background functions are used in the minimization of the negative logarithm of the likelihood with a penalty term added to account for the number of free parameters in the background model. The profile likelihood ratio for the penalized likelihood function can be written as
\begin{linenomath}
\begin{equation}
\label{eq:ll}
-2 \ln \frac{\widetilde{\mathcal{L}}({\rm data}|\mu, \hat\theta_\mu, \hat b_\mu)}{\widetilde{\mathcal{L}}({\rm data}|\hat\mu, \hat\theta, \hat b)}.
\end{equation}
\end{linenomath}
In this equation the numerator is the maximum penalized likelihood for a given $\mu$, at the best-fit values of nuisance parameters, $\hat\theta_\mu$, and of the background function, $ \hat b_\mu$. The denominator is the global maximum for $\widetilde{\mathcal{L}}$, achieved at $\mu=\hat\mu$, $\theta = \hat\theta$, and $b = \hat b$. A confidence interval on $\mu$ is obtained with the background function maximizing $\widetilde{\mathcal{L}}$ for any value of $\mu$~\cite{Dauncey:2014xga}.

\begin{figure}[!ht]
\begin{center}
\includegraphics[angle=0,width=0.45\textwidth]{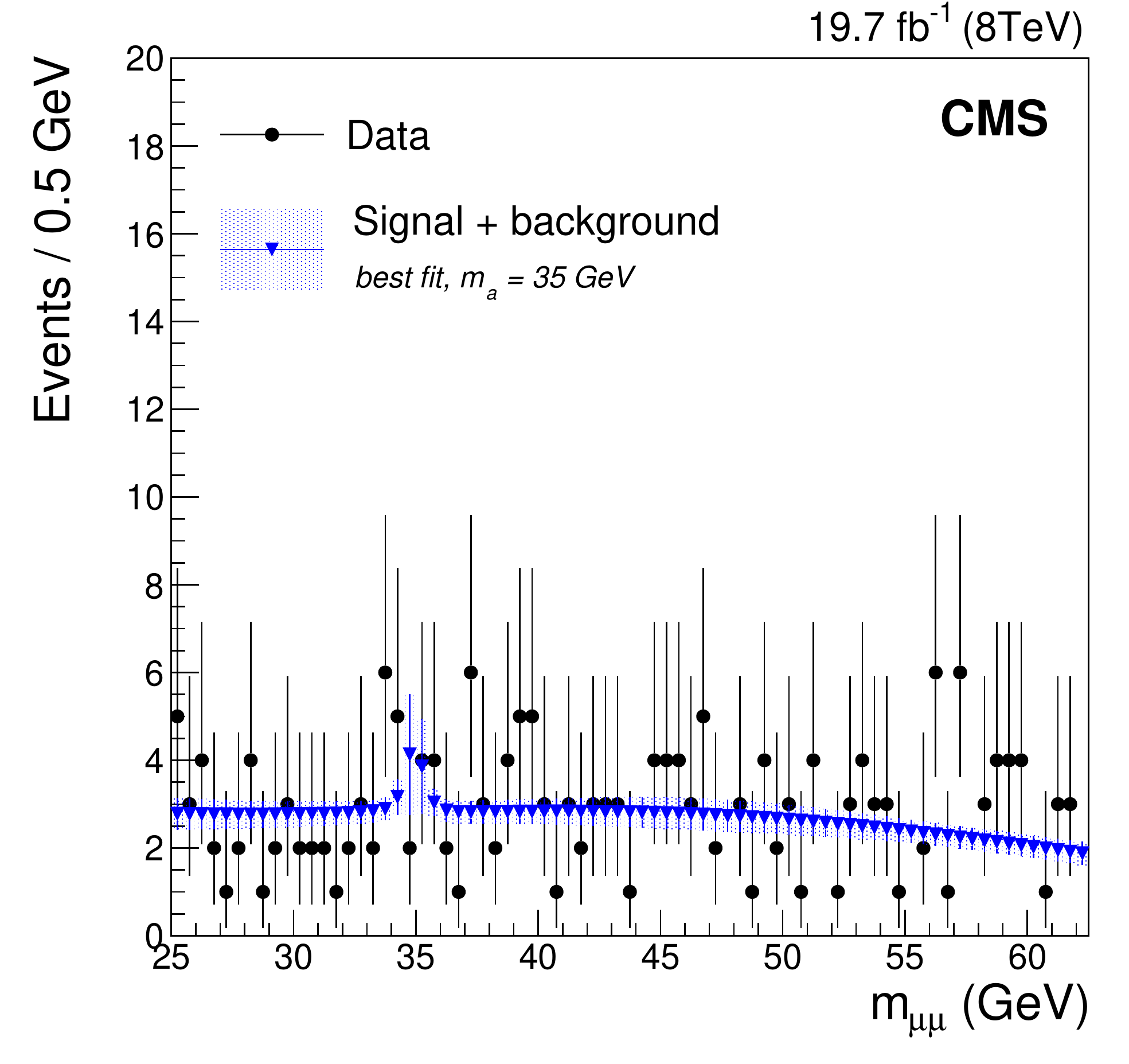}  \\
\caption{The best fit to the data for a signal-plus-background model with $\ma=35\GeV$, including profiling of the uncertainties, in the search for $\processmmbb$ events.}
\label{fig:postfit_mmbb}
\end{center}
\end{figure}

The analysis of data yields no significant excess of events over the SM background prediction. Figure~\ref{fig:postfit_mmbb} shows the $\mmumu$ distribution in data together with the best fit output for a signal-plus-background model at $\ma=35\GeV$. The relative difference between the expected limit of the best-fit background model and that of the unconditional fit is about 40\%.

\section{Search for \texorpdfstring{$\processmmtt$}{mmtt} decays}
\label{sec:mmtt}

Five final states are studied in the $\processmmtt$ channel,
depending on whether the $\PGt$ leptons decay to electrons ($\PGt_{\Pe}$),
to muons ($\PGt_{\PGm}$), or to hadrons ($\PGt_{\text{h}}$): $\PGm^+\PGm^-\PGt_{\Pe}^+\PGt_{\Pe}^-$, $\PGm^+\PGm^-\PGt_{\Pe}^\pm\PGt_{\PGm}^\mp$, $\PGm^+\PGm^-\PGt_{\Pe}^\pm{\PGt_{\text{h}}}^\mp$,
$\PGm^+\PGm^-\PGt_{\PGm}^\pm{\PGt_{\text{h}}}^\mp$, or $\PGm^+\PGm^-{\PGt_{\text{h}}}^+{\PGt_{\text{h}}}^-$.
The $\PGm^+\PGm^-\PGt_{\PGm}^+\PGt_{\PGm}^-$ final state is not considered due to the difficulty of correctly identifying the reconstructed muons as either direct pseudoscalar or $\PGt$ decay products, which results in low sensitivity.
Given the 2\% dimuon mass resolution for the muons originating promptly from one of the $\ab$ bosons arising from the $\Hb$ boson decay, an unbinned likelihood fit is performed to extract the results, using
$\mmumu$ as the observable. Pseudoscalar boson masses between 15 and 62.5\GeV are probed; the lower bound corresponds to the minimum mass that ensures a good signal efficiency with selection criteria that do not rely on boosted lepton pairs, and an expected background large enough to be modeled through techniques described below.

\subsection{Event selection}

Events are selected using a double muon trigger relying on the presence of a muon with $p_{\rm T}>17\,\GeV$ and another one with $p_{\rm T}>8\,\GeV$. For the offline selection, the leading muon $p_{\rm T}$ threshold is increased to 18 $\GeV$, while the subleading muon $p_{\rm T}$ must exceed 9 $\GeV$.
To reconstruct the dimuon pair from the $\ab\to\PGm^+\PGm^-$ decay, two isolated muons of opposite charge,
$\pt> 5\GeV$, and $|\eta|< 2.4$ are selected.
In the $\PGm^+\PGm^-\PGt_{\Pe}^+\PGt_{\Pe}^-$, $\PGm^+\PGm^-\PGt_{\Pe}^\pm{\PGt_{\text{h}}}^\mp$ and
$\PGm^+\PGm^-{\PGt_{\text{h}}}^\pm{\PGt_{\text{h}}}^\mp$ final states, where these are the only muons, their $\pt$ thresholds are raised to 18 and 9\GeV to match the trigger requirements.
If there are more than two muons in the final state,
the highest-$\pt$ muon is required to pass a $\pt$ threshold of 18\GeV, and is considered as arising from the prompt decay of the light boson.
It is then paired with the next highest-$\pt$ muon of opposite charge. The other muons are considered to arise from leptonic decays of the $\PGt$ lepton. The second highest-$\pt$ muon is required to have $\pt$ greater than 9\GeV.
Muons are paired correctly in about 90\% of the events for all masses.
The $\PGt\PGt$ pair is reconstructed from a combination of oppositely charged identified and isolated muons,
electrons, or $\PGt_{\text{h}}$, depending on the final state.
The muons are selected with $\pt>5\GeV$ and $|\eta|<2.4$, the electrons with $\pt>7\GeV$ and $|\eta|<2.5$, and the $\PGt_{\text{h}}$ candidates
with $\pt>15\GeV$ and $|\eta|<2.3$. The contribution from $\Hb\to \PZ\PZ^*\to \PGm^+\PGm^- \Pe^+\Pe^-$ events is suppressed, in the $\PGm^+\PGm^- \Pe^+\Pe^-$ final state, by
excluding events with visible invariant mass of the four leptons inside a 30\GeV-wide window around 125\GeV, the Higgs boson mass. The signal efficiency of this selection criterion is high since the four lepton invariant mass in $\Pgm^+\Pgm^-\Pgt_e^+\Pgt_e^-$ events is significantly reduced due to the presence of neutrinos in $\Pgt$ lepton decays.

The four objects are required to be separated from each other by at least $\Delta R$ = 0.4.
Events are discarded if they contain at least one jet that satisfies a b-tag requirement that allows $\mathcal{O}$(0.1\%) of the light quark jets to survive, while the tag efficiency for genuine b jets is about 50\%. This reduces the contribution from backgrounds with top quarks.
To prevent a single event from contributing to different final states, events containing
other identified and isolated electrons or muons in addition to the four selected objects are rejected; less than 1\% of signal events are rejected because of this veto.
Two selection criteria with a high signal efficiency are designed to reduce the contribution of the backgrounds to the signal region:
the invariant mass of the $\PGm\PGm\PGt\PGt$ system is required to lie close to the Higgs boson mass
($|m_{\PGm\PGm\PGt\PGt}-125\GeV|<$ 25\GeV), and the normalized difference between the masses of the di-$\PGt$ and dimuon systems is required to be small
($|\mmumu -m_{\PGt\PGt}|/\mmumu <$ 0.8). The $\PGt\PGt$ mass, $m_{\PGt\PGt}$, used to define both variables,
is fully reconstructed with a maximum likelihood algorithm
taking as input the four-momenta of the visible particles, as well as the $\Etmiss$ and its resolution~\cite{Bianchini:2014vza}.
This method gives a resolution of about 20\% and 10\%, for the $\PGt\PGt$ mass $m_{\PGt\PGt}$ and four-lepton mass $m_{\PGm\PGm\PGt\PGt}$, respectively. Finally, only events with a reconstructed dimuon mass between 14 and 66\GeV are considered in the study.

\subsection{Signal and background estimation}

Two types of backgrounds contribute to the signal region: irreducible ZZ production, and reducible processes with
at least one jet being misidentified as one of the final-state leptons, mainly composed of Z+jets and WZ+jets events.
The $\PZ\PZ\to4\ell$ contribution, where $\ell$ denotes any charged lepton, is estimated from MC simulations, and the process is
scaled to the NLO cross section~\cite{ZZNLO}.
The normalization and $\mmumu$ distribution of the reducible processes are determined separately, using control samples in data.
To estimate the normalization, the rates for jets to be misidentified as
$\PGt_{\text{h}}$, electrons, or muons are measured in dedicated signal-free control regions, defined similarly to the
signal region except that the $\PGt$ candidates (electrons, muons, or $\PGt_{\text{h}}$)
pass relaxed identification and isolation conditions and have SS charge.
All misidentification rates are measured as a function of the \pt of
the jet closest to the $\PGt$ candidate, and are fitted using a decreasing exponential in addition to a constant term.
Events with $\PGt$ candidates passing the relaxed identification and isolation conditions, but not the signal region criteria, are scaled with weights
that depend on the misidentification rates, to
obtain an estimate
of the yield of the reducible background in the signal region.
The $\mmumu$ distribution of reducible backgrounds is taken from a signal-free region in data,
where both $\PGt$ candidates have SS charge and pass relaxed identification and isolation criteria.

The dimuon mass distribution in signal events in final states with two muons is parameterized with a Voigt profile.
In final states with three muons, the Gaussian component of the profile is found to be negligible, and the signal distributions are parameterized with Breit--Wigner profiles.
A fit is performed for every final state and every generated $\ab$.
To interpolate the signal distributions to any $\ab$ boson in the studied mass range, the parameters of the fit functions
are parameterized as a function of $\ma$ by fitting with a third-degree
polynomial the parameters of the Voigt or Breit--Wigner profiles obtained from the individual fits.
A similar technique is used to interpolate the signal normalization to
intermediate mass points; the parameterization leads to yield uncertainties for the signal between 5 and 8\% depending on the final state.
A closure test that consists of removing a signal sample corresponding to a given mass point from the parameterization of the Voigt and Lorentz fit parameters as a function of the mass,
then comparing the parameterization interpolation
to the direct fit to this sample, has demonstrated the validity of this technique.
The ZZ irreducible background and reducible backgrounds are parameterized with Bernstein polynomials with five and three degrees of freedom respectively.
The degrees of the polynomials are chosen to be the lowest that allow for a good agreement between the fit functions
and the predicted backgrounds, according to f-tests. Uncertainties in the fit parameters of the Bernstein polynomials for reducible processes are taken into account in the
statistical interpretation of results. They dominate over uncertainties associated with the
choice of the fitting functions, which are neglected. Uncertainties in the ZZ background distribution are neglected given the low expected yield for this process
relative to the reducible background contribution.

\begin{figure}[htbp]
\begin{center}
\includegraphics[angle=0,width=0.65\textwidth]{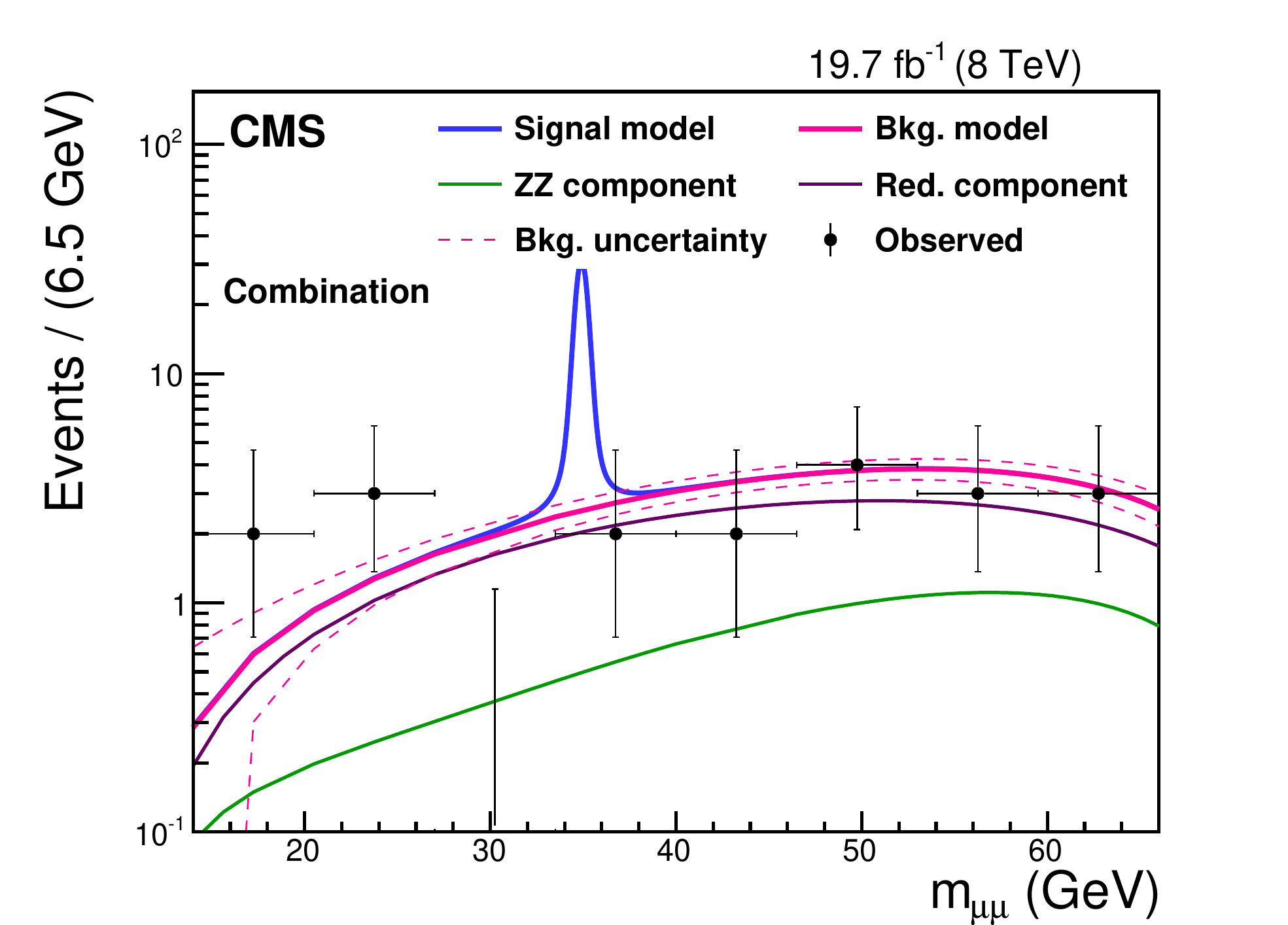}
\caption{Background and signal ($\ma= 35\GeV$) models, scaled to their expected yields, for the combination of all final states ($\PGm^+\PGm^-\PGt_{\Pe}^+\PGt_{\Pe}^-$, $\PGm^+\PGm^-\PGt_{\Pe}^\pm\PGt_{\PGm}^\mp$, $\PGm^+\PGm^-\PGt_{\Pe}^\pm{\PGt_{\text{h}}}^\mp$,
$\PGm^+\PGm^-\PGt_{\PGm}^\pm{\PGt_{\text{h}}}^\mp$, and $\PGm^+\PGm^-{\PGt_{\text{h}}}^+{\PGt_{\text{h}}}^-$) in the search for $\processmmtt$ decays.
The two components of the background model, ZZ and reducible processes, are drawn.
The signal sample is
scaled with $\sigma_{\Hb}$ as predicted in the SM, assuming $\mathcal{B}(\Hb\to \ab\ab) = 10\%$, and considering decays of the pseudoscalar
$\ab$ boson to leptons only ($\mathcal{B}(\ab\to{\PGt^{+}}{\PGt^{-}})+\mathcal{B}(\ab\to\PGm^+\PGm^-)+\mathcal{B}(\ab\to \Pe^+\Pe^-) = 1$) using Eq.~(\ref{eq:2hdm}).
The results are shown after a simultaneous maximum likelihood fit in all five channels that takes into account the systematic uncertainties described in Section~\ref{sec:systematics}.
}
\label{fig:massplots}
\end{center}
\end{figure}

The parameterized dimuon mass distributions and the observed events after the complete selections are shown in Fig.~\ref{fig:massplots}
for the combination of the five final states.
In this figure, the signal sample is normalized based on the Higgs boson cross section, $\sigma_{\Hb}$, predicted in the SM. A branching fraction of 10\% is assumed for $\Hb\to \ab\ab$. The a boson is assumed to decay only to leptons ($\mathcal{B}(\ab\to{\PGt^{+}}{\PGt^{-}})+\mathcal{B}(\ab\to\PGm^+\PGm^-)+\mathcal{B}(\ab\to \Pe^+\Pe^-) = 1$), using Eq.~(\ref{eq:2hdm}).
Combining all final states, 19 events are observed while $20.7\pm 2.2$ are expected in the absence of signal. The expected signal yield, assuming the normalization  described
above, ranges from 3.1 to 8.2 events over the probed mass range, as detailed in Table~\ref{tab:yieldsmmtt}.
\begin{table}
\renewcommand{\arraystretch}{1.1}
\begin{center}
\caption{Expected and observed yields in the search for $\processmmtt$ decays. The signal samples are
scaled with the production cross section for the SM $\Hb$ boson, assuming $\mathcal{B}(\Hb\to \ab\ab) = 10\%$ and considering decays of the pseudoscalar
$\ab$ boson to leptons only. Background yields are obtained after a maximum likelihood fit to observed data, taking into account the
systematic uncertainties detailed in Section 6.}
\begin{tabular}{c|cc|ccc|c}
& \multicolumn{2}{c|}{Signal} & \multicolumn{3}{c|}{Backgrounds} & \multirow{2}{*}{Obs.} \\
& $\ma = 20\GeV$ & $\ma = 60\GeV$ & ZZ & Reducible & Total & \\
\hline
$\PGm^+\PGm^-\PGt_{\Pe}^+\PGt_{\Pe}^-$ & 0.20$\pm$0.02 &  0.58$\pm$0.06 & 4.71$\pm$0.47 & 2.56$\pm$1.06 & 7.27$\pm$1.16& 8 \\
$\PGm^+\PGm^-\PGt_{\Pe}^\pm\PGt_{\PGm}^\mp$ & 0.58$\pm$0.08 & 1.42$\pm$0.16 & 0.10$\pm$0.01 & 1.68$\pm$0.70 & 1.78$\pm$0.70& 2 \\
$\PGm^+\PGm^-\PGt_{\Pe}^\pm{\PGt_{\text{h}}}^\mp$ & 0.74$\pm$0.08 & 2.02$\pm$0.20 & 0.16$\pm$0.02 & 5.66$\pm$1.48 & 5.82$\pm$1.48 & 5 \\
$\PGm^+\PGm^-\PGt_{\mu}^\pm{\PGt_{\text{h}}}^\mp$ & 0.96$\pm$0.10 & 2.30$\pm$0.22 & 0.13$\pm$0.02 & 0.91$\pm$0.28 & 1.14$\pm$0.29 & 1 \\
$\PGm^+\PGm^-{\PGt_{\text{h}}}^+{\PGt_{\text{h}}}^-$ & 0.60$\pm$0.06  & 1.90$\pm$0.18 & 0.06$\pm$0.02 & 4.64$\pm$0.94 & 4.70$\pm$0.94 & 3 \\
\hline
Combined & 3.08$\pm$0.31 & 8.22$\pm$0.82 & 5.09$\pm$0.39 & 15.47$\pm$2.41\phantom{0} & 20.71$\pm$2.23\phantom{0} &  19 \\
\end{tabular}
\label{tab:yieldsmmtt}
\end{center}
\end{table}

\section{Systematic uncertainties}
\label{sec:systematics}

The statistical interpretation of the analyses takes into account several sources of systematic uncertainties,
included in the likelihood function as nuisance parameters following log-normal distributions in the case of yield uncertainties. Uncertainties related
to the modeling of backgrounds estimated from data have already been discussed for the three independent analyses
in Sections~\ref{sec:tttt},~\ref{sec:mmbb}, and~\ref{sec:mmtt}, and will only be partially described here. Other systematic uncertainties are detailed in the following subsections, and summarized in Table~\ref{tab:sys}.
\subsection{Systematic uncertainties common to all analyses}
Systematic uncertainties common to all analyses include the uncertainties in the trigger efficiency (between 0.2 and 4.2\% depending
on the analysis and on the process), the lepton identification and isolation efficiencies (6\% for every $\PGt_{\text{h}}$~\cite{TauID}, between 0.5 and 1.5\% for
muons, 2\% for electrons), all evaluated with tag-and-probe methods~\cite{tagandprobe} in Drell-Yan data and simulated samples.
The uncertainties associated with the data-to-simulation correction factor for the $\PQb$ tagging efficiencies and misidentification rates
are also propagated as systematic uncertainties to the final results ~\cite{1748-0221-8-04-P04013}.
Uncertainties in the knowledge of the parton distribution functions~\cite{Alekhin:2011sk,Botje:2011sn} are taken into account as yield uncertainties, and do not
affect the shape of signal mass distributions.
The uncertainty in the integrated luminosity amounts to 2.6\%.
\subsection{Systematic uncertainties for the \texorpdfstring{$\processtttt$}{tttt} search}
The leading systematic uncertainty in the $\processtttt$ analysis comes from imperfect knowledge of the background composition in the signal region; it amounts to up to 90\% of the background yield, as discussed in Section~\ref{sec:tttt}.
Other sources of systematic uncertainty specific to this search affect the expected signal yield only.  When added in quadrature to the background uncertainty, signal yield uncertainties account for at most 6 (10)\% of the total uncertainty for $\mt \leq$ ($>$) 50\GeV.  These minor uncertainties include an additional uncertainty of up to 10\% related to
the muon isolation if the trigger muon comes from a boosted $\PGt_{\PGm}\PGt_{\text{X}}$ topology, as in the $ \cPg\cPg\Hb $, $ \PZ\Hb $, and VBF production modes, rather than an isolated $\PW$ leptonic decay, as in the $ \PW\Hb $ mode.
The signal yield is further affected by an asymmetric uncertainty in the $\PGt$ charge misidentification probability of ${-}1\%$ and ${+}2\%$. Up to 9.3\%
uncertainty in the signal yield is considered to account for uncertainties in the $\mt$ computation because of uncertainties in the \Etmiss measurements. The $\PQb$ veto on
the seed jet of the $\PGt_{\text{X}}$ object introduces a maximum of 9.4\% uncertainty in the signal yield. Finally, it should be noted that the full MC simulation and event reconstruction were only performed for the $ \cPg\cPg\Hb $ and $ \PW\Hb $ samples with $m_{\ab}$ = 5, 7, 9, 11, 13, and 15\GeV, and for the VBF and $ \PZ\Hb $ samples with $m_{\ab}$ = 9\GeV.  The yields for the VBF ($ \PZ\Hb $) samples with $m_{\ab}$ = 5, 7, 11, 13, and 15\GeV were extrapolated from the $ \cPg\cPg\Hb $($ \PW\Hb $) simulated samples at the corresponding pseudoscalar mass, which have similar final state kinematics.  An uncertainty between 19\% and 25\%, depending on the production mode and $\mt$ bin, is assigned to cover imperfect knowledge of the acceptance for the signals that were not simulated.

\subsection{Systematic uncertainties for the \texorpdfstring{$\processmmbb$}{mmbb} search}

For the $\processmmbb$ analysis, the energy of jets is varied within a set of uncertainties depending on the jet $\pt$ and $\eta$. This amounts to a 7\% variation of the expected signal yield. The jet smearing corrections are altered within their uncertainties~\cite{CMSJetPaper} to account for the uncertainty arising from the jet energy resolution, which has an effect on the process yield of about 1\%. Furthermore, the uncertainty in the amount of pileup interactions per event is estimated by varying the total inelastic pp cross section~\cite{tagkey20135} by ${\pm}5\%$. All sources of uncertainties including those associated with the muon energy scale and reconstruction and identification efficiencies are found to have a negligible effect on the signal modeling. The signal shape parameters are therefore left floating within their statistical uncertainties in the fit. The systematic uncertainty related to the discrete profiling method is small compared to the statistical uncertainty.

\subsection{Systematic uncertainties for the \texorpdfstring{$\processmmtt$}{mmtt} search}

The effect of the $\PGt_{\text{h}}$ energy scale in the $\processmmtt$ analysis is propagated to the mass distributions, and leads to uncertainties in the yields of the signal and of the irreducible background between
0 and 10\%, depending on the final state.
The muon energy scale uncertainty, amounting to 0.2\%, is found to shift the mean of the signal distributions by up to 0.2\%; this is taken
into account as a parametric uncertainty in the mean of the signal distributions.
Statistical uncertainties in the parameterization of the signal are accounted for through the uncertainties on the fit parameters describing the signal shape.
The uncertainty in the normalization of the reducible background is obtained by varying the fit functions of the misidentification rates
within their uncertainties. Uncertainties in background yields lie between 25 and 50\%; uncertainties related to a given misidentification rate are correlated
between corresponding final states.
The number of events in the MC simulation of the ZZ background passing the full signal selection is small, and a statistical uncertainty ranging between
1 and 15\% depending on the final state is considered to take this effect into account.
This uncertainty is uncorrelated across all final states.
\begin{table}
\renewcommand{\arraystretch}{1.1}
\begin{center}
\topcaption{Sources of systematic uncertainties, and their effects on process yields, for the three different searches.
}\label{tab:sys}
\begin{tabular}{l|c|c|c}
\multirow{2}{*}{Source of uncertainty}  & \multicolumn{3}{c}{Uncertainty in acceptance (\%)} \\
& $4\PGt$ & $2\PGm2\PQb$ & $2\PGm2\PGt$ \\
\hline
Luminosity & 2.6 & 2.6 & 2.6 \\
Trigger efficiency & 0.2-4.2 & 1.5 & 1 \\
$e$ identification & 1 & --- & 0-4 \\
$\PGm$ identification & 0.5-1.5 & 3.5 & 2-3 \\
+ for boosted $\tau_\mu\tau_X$ objects & 10 & --- & --- \\
$\tau_h$ identification & 6 & --- & 0-12 \\
$\PQb$ tagging & 0.2-9.4 & 0.1-4.5 & 1 \\
Data-driven background estimation & 59-84 & discrete profiling & 25-50 + shape unc. \\
Tau charge misidentification & 2 & --- & --- \\
\Etmiss scale & 1-9 & --- & --- \\
VBF and $\PZ\Hb$ extrapolation & 19-25 & --- & --- \\
Jet energy scale & --- & 7 & --- \\
Jet energy resolution & --- & 0.10-0.15 & --- \\
Tau energy scale & --- & --- & 0-10 \\
Muon energy scale & --- & 3.5 & Shape unc. only \\
ZZ simulation size & --- & --- & 1-15 \\
ZZ cross section & --- & --- & 5+6 \\
\end{tabular}
\end{center}
\end{table}

\section{Results}
\label{sec:results}

\subsection{Results of the search for \texorpdfstring{$\processtttt$}{tttt} decays}

\begin{figure}[!hbtp]
\begin{center}
\includegraphics[width=1.15\cmsFigWidth]{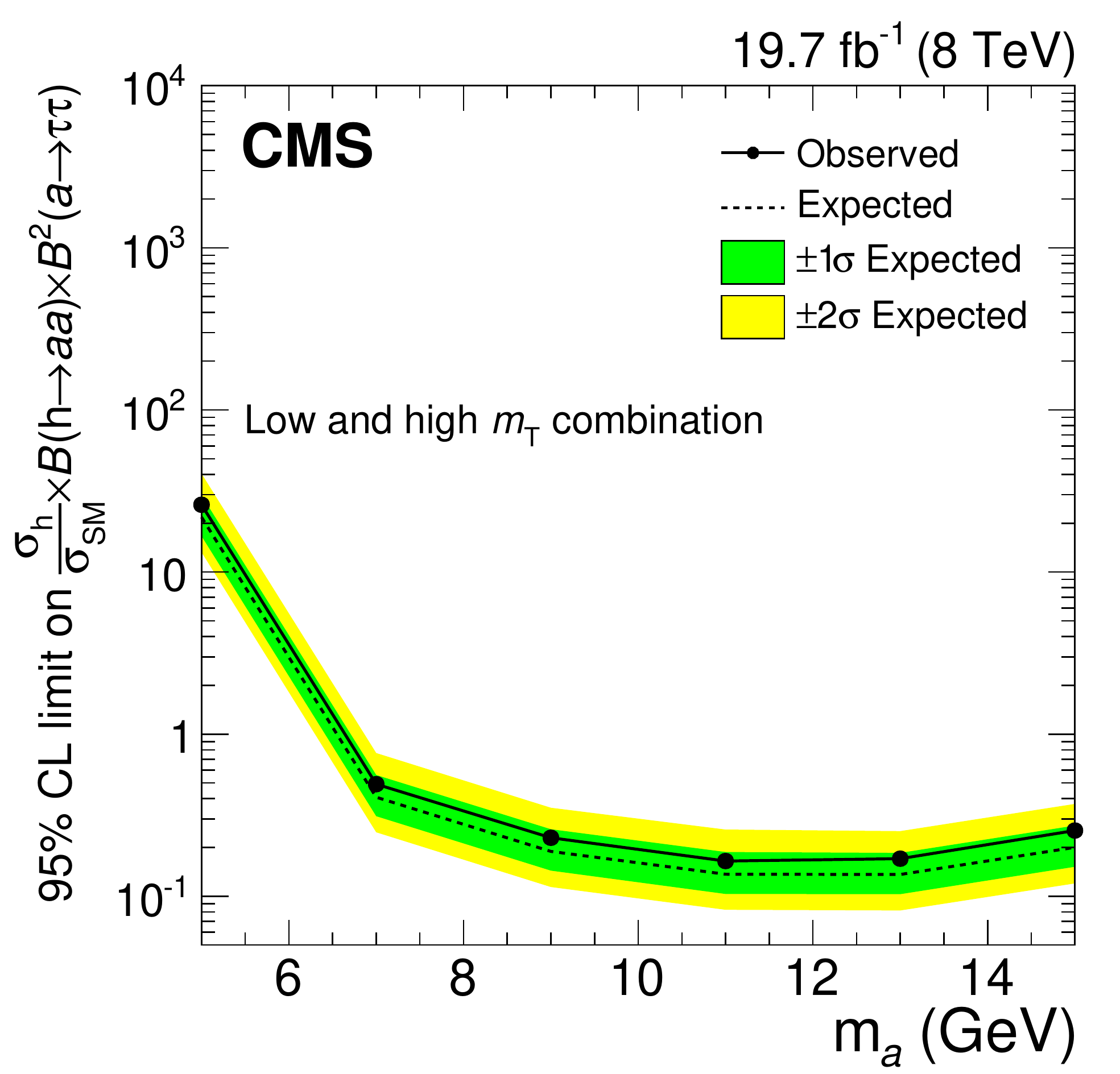}
\caption{Observed 95\% CL limits on the branching fraction $\mathcal{B}(\Hb\to \ab\ab) \, \mathcal{B}^{2}(\ab \to \PGt^{+}\PGt^{-})$ assuming SM \Hb production rates, compared to expected limits for pseudoscalar mass points between 5 and 15\GeV.}
\label{fig:lowhighMTCLs}
\end{center}
\end{figure}

The number of events observed in the signal window is compatible with the SM background prediction for the $\Hb\to\ab\ab\to4\PGt$ analysis.
Results are interpreted as upper limits on the production of $\Hb\to aa$ relative to the SM Higgs boson production, scaled by
$\mathcal{B}(\Hb\to \ab\ab) \, \mathcal{B}^{2}(\ab \to \PGt^{+}\PGt^{-}) \equiv \mathcal{B}(\processtttt)$. SM production cross sections are taken for $ \cPg\cPg\Hb $, $ \PW\Hb $, $ \PZ\Hb $, and VBF processes~\cite{LHCHXSWG}. Upper limits are calculated using the
modified $\text{CL}_{\text{s}}$ technique~\cite{CMS-NOTE-2011-005,Cowan:2010js,Junk,Read},
in which the test statistic is a profile likelihood ratio. The asymptotic approximation is used to extract the results.  In Figures~\ref{fig:lowhighMTCLs},~\ref{fig:postfit}, and~\ref{fig:limits_mmtt}, the green (yellow) band labeled ``${\pm}1(2)\sigma$ Expected'' denotes the expected 68 (95)\% C.L. band around the median upper limit if no data consistent with the signal expectation were to be observed.

The expected limits and the observed limit for the combination of the low- and high-$\mt$
bin as a function of $\ma$ are shown in Fig.~\ref{fig:lowhighMTCLs}.  The sharp decrease in sensitivity between 5 and 7\GeV results from the 4\GeV $m_{\PGm+\text{X}}$ signal requirement, which is less efficient for lower mass pseudoscalars.

\subsection{Results of the search for \texorpdfstring{$\processmmbb$}{mmbb} decays}

\begin{figure}[!htbp]
\begin{center}
\includegraphics[angle=00,width=0.45\textwidth]{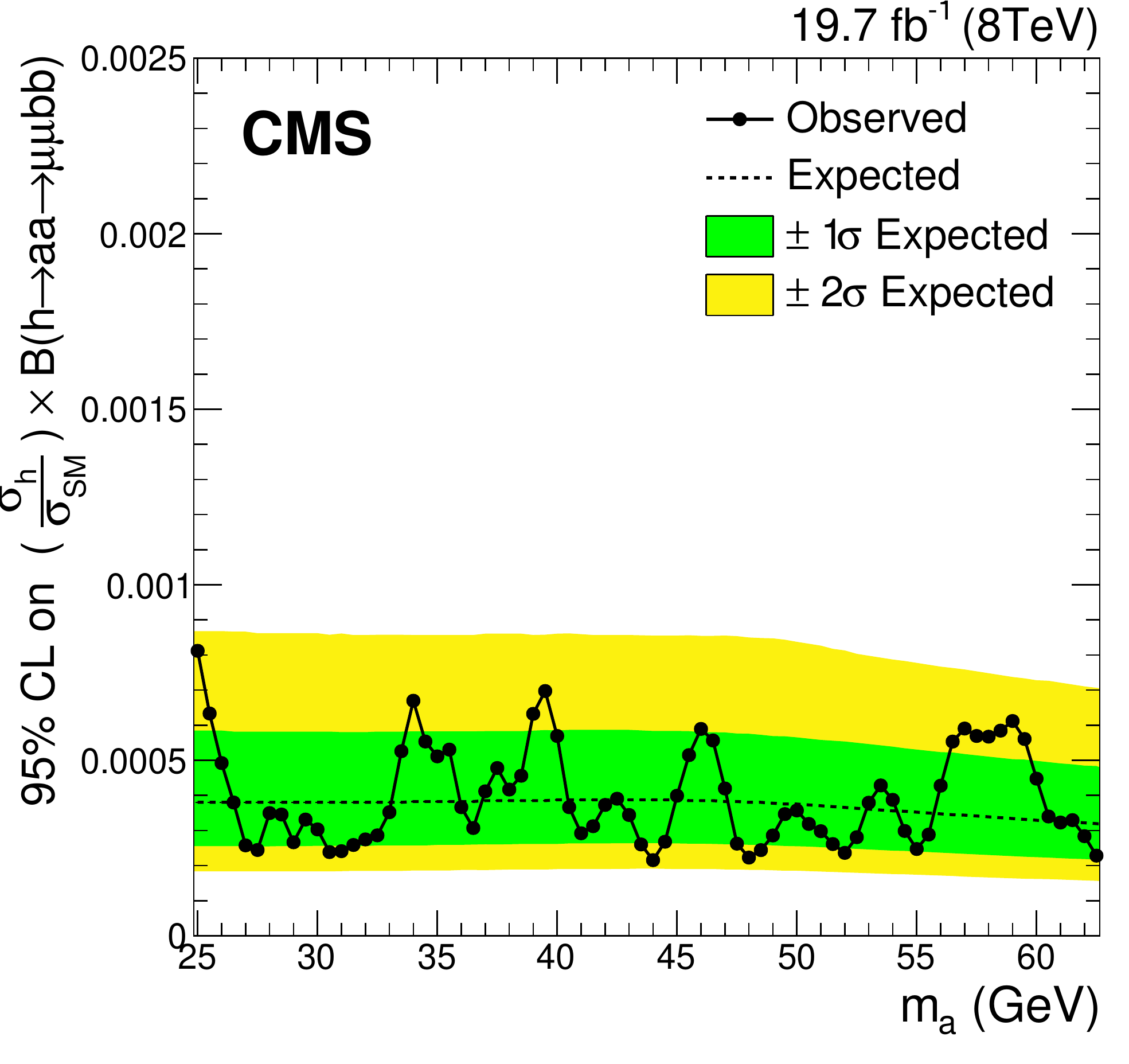}
\caption{Observed and expected upper limits at 95\% CL on the $\Hb$ boson production normalized to the SM prediction times $\mathcal{B}(\processmmbb)$.}
\label{fig:postfit}
\end{center}
\end{figure}

The analysis of the mass spectrum for the $\processmmbb$ search does not show any significant excess of events over the SM background prediction either,
as seen in Fig.~\ref{fig:postfit_mmbb}. Upper limits on the production of $\Hb\to\ab\ab$ relative to
the SM Higgs boson $ \cPg\cPg\Hb $ production mode,
scaled by $\mathcal{B}({\ab}\to{\PQb}\overline{\PQb})\, \mathcal{B}({\ab}\to \PGm^+ \PGm^-)$, are obtained at 95\% CL with the asymptotic $\text{CL}_{\text{s}}$ method.
The observed and expected limits, together with the expected uncertainty bands, are illustrated in Fig.~\ref{fig:postfit}.
The oscillations in the observed limit arise from the narrow dimuon mass resolution predicted for signal events.

\subsection{Results of the search for \texorpdfstring{$\processmmtt$}{mmtt} decays}

\begin{figure}[htbp!]
\begin{center}
\includegraphics[width=0.48\textwidth]{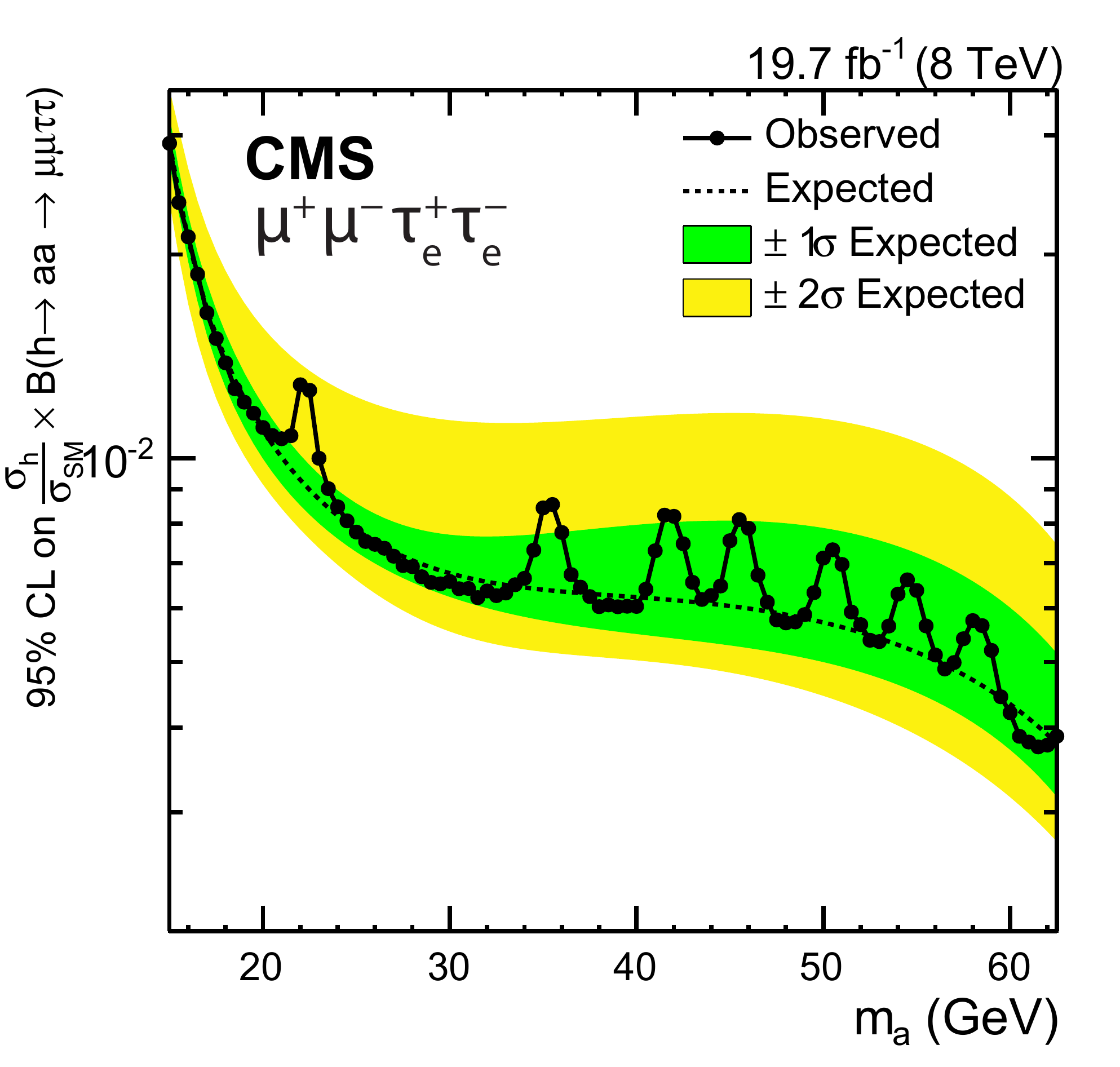}
\includegraphics[width=0.48\textwidth]{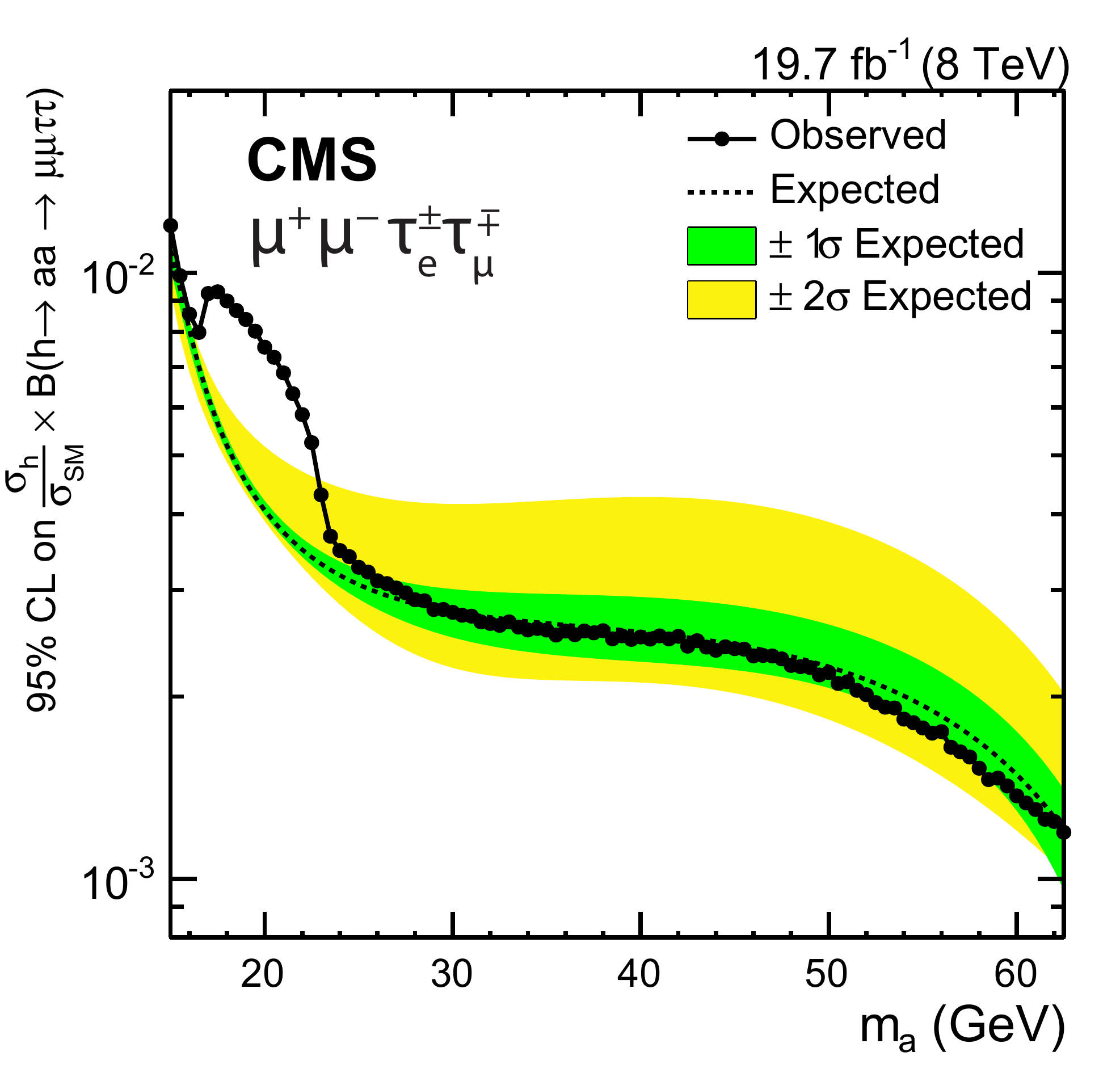}
\includegraphics[width=0.48\textwidth]{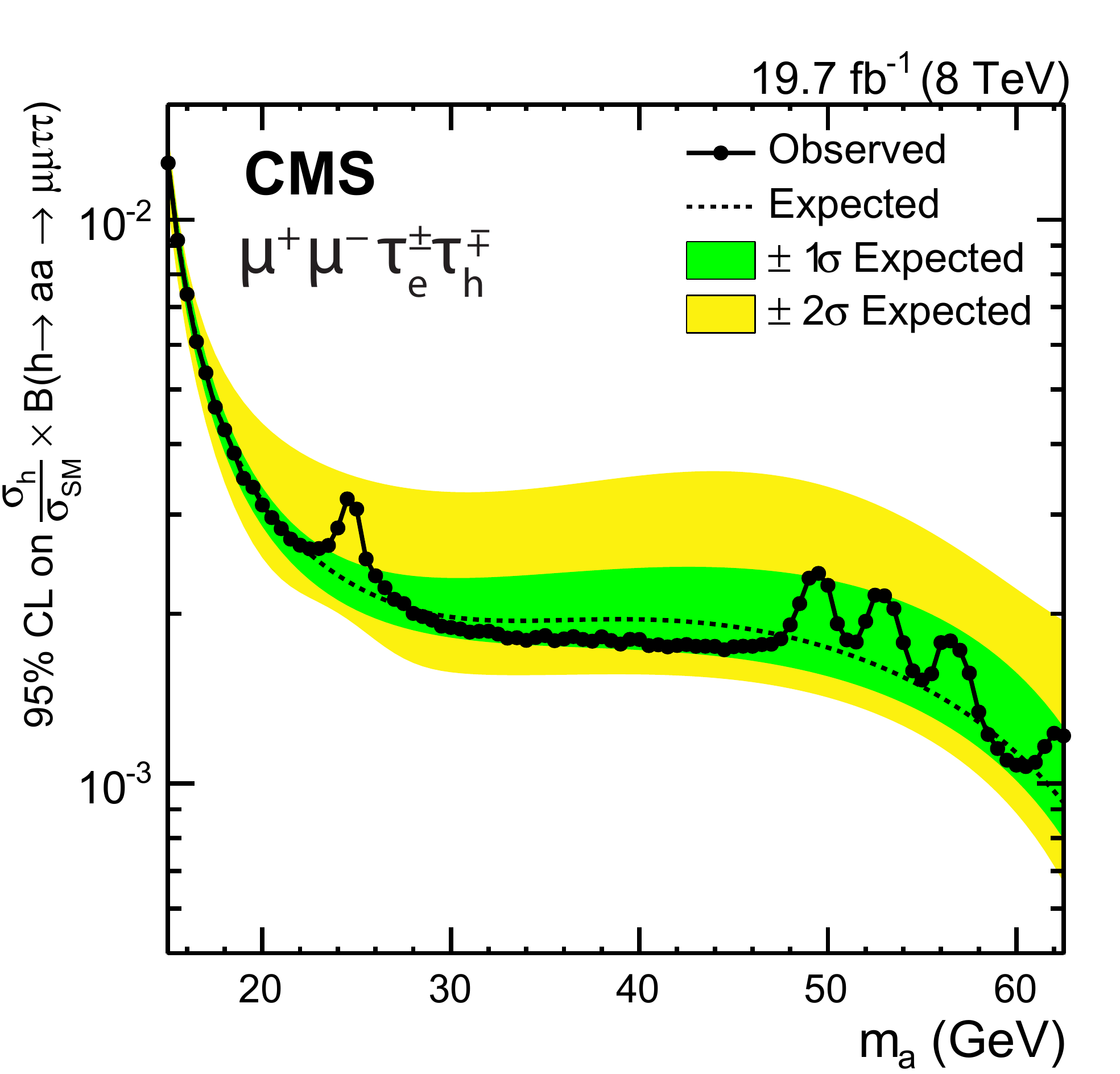}
\includegraphics[width=0.48\textwidth]{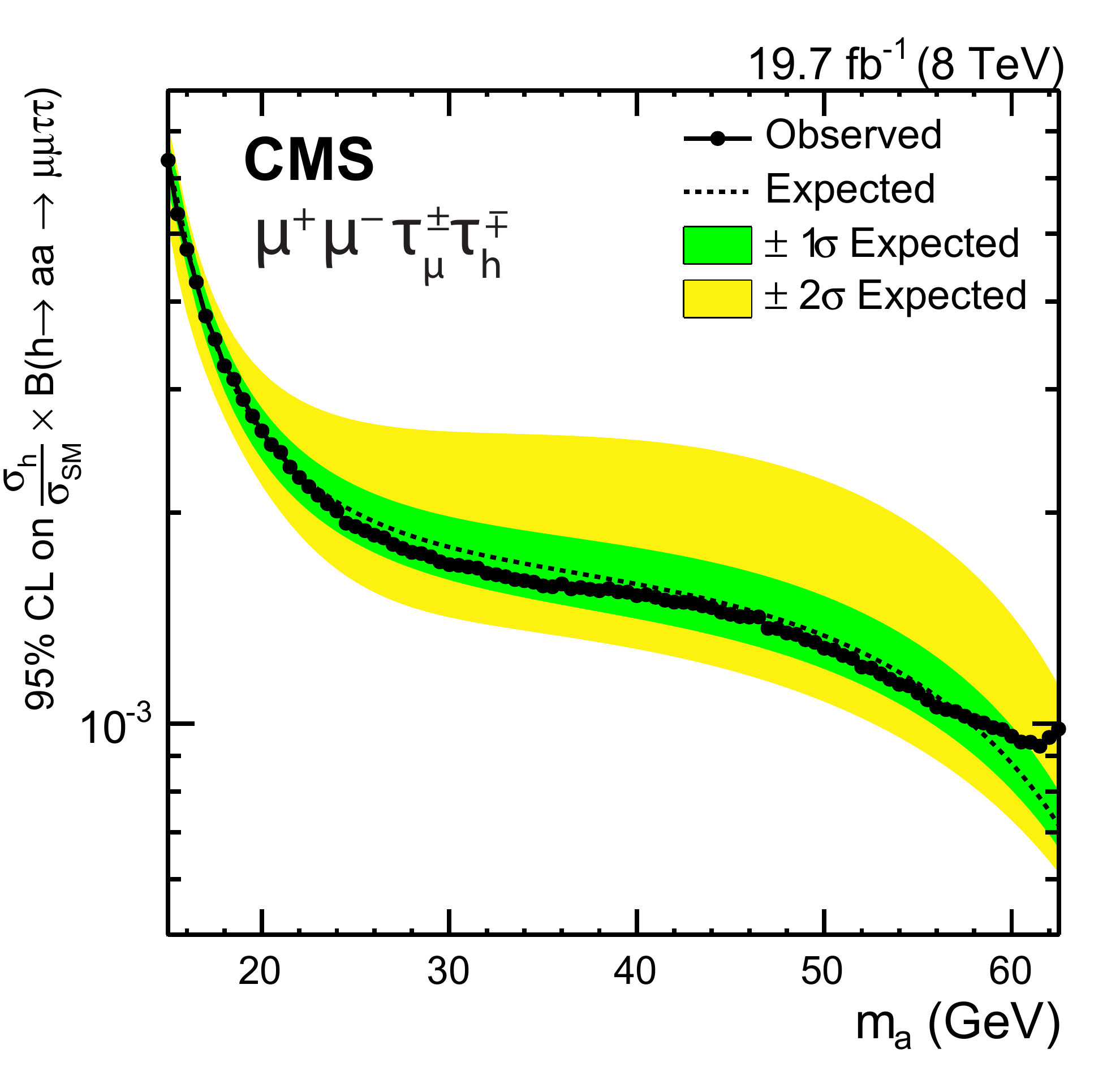}
\includegraphics[width=0.48\textwidth]{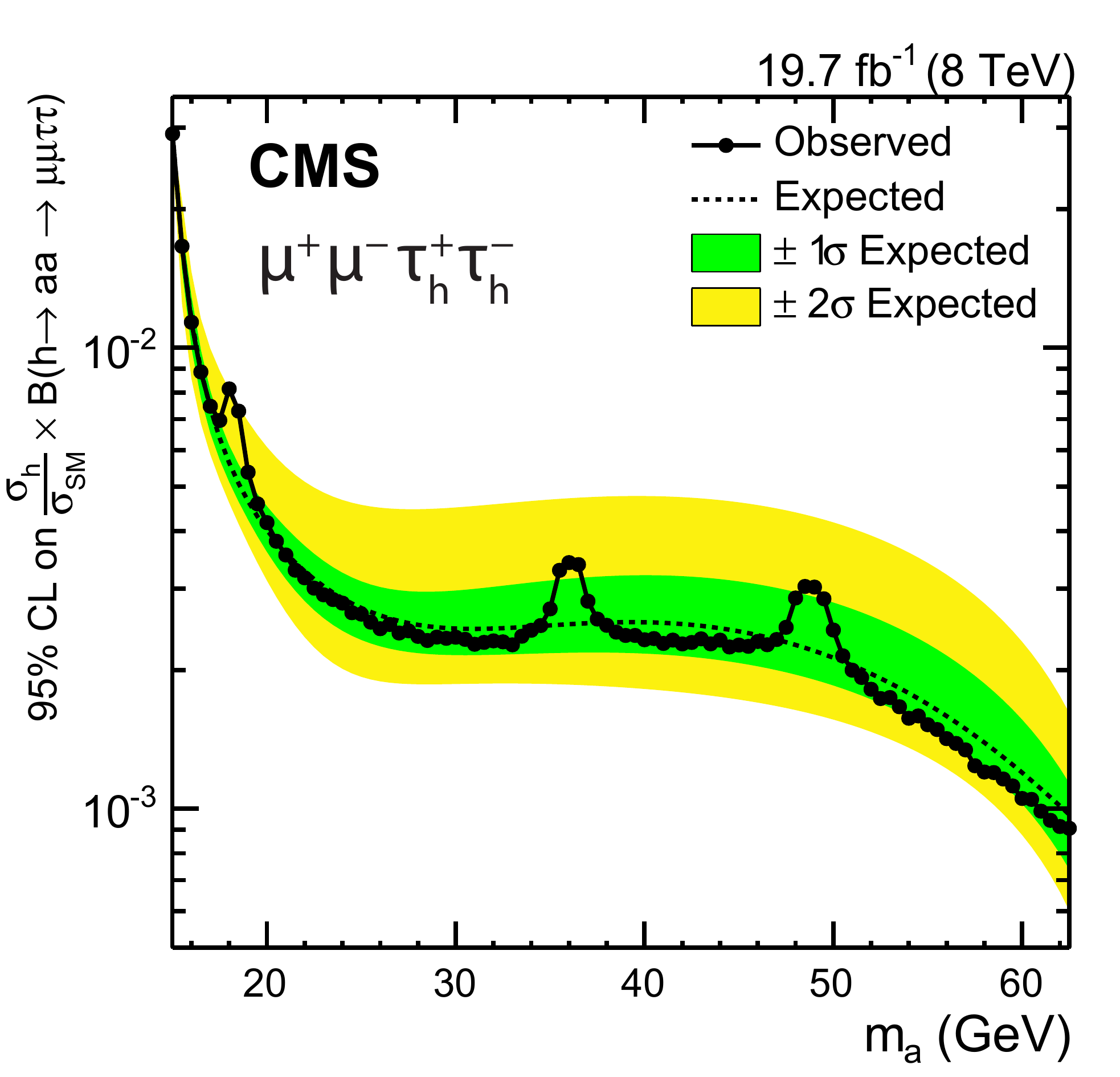}
\includegraphics[width=0.48\textwidth]{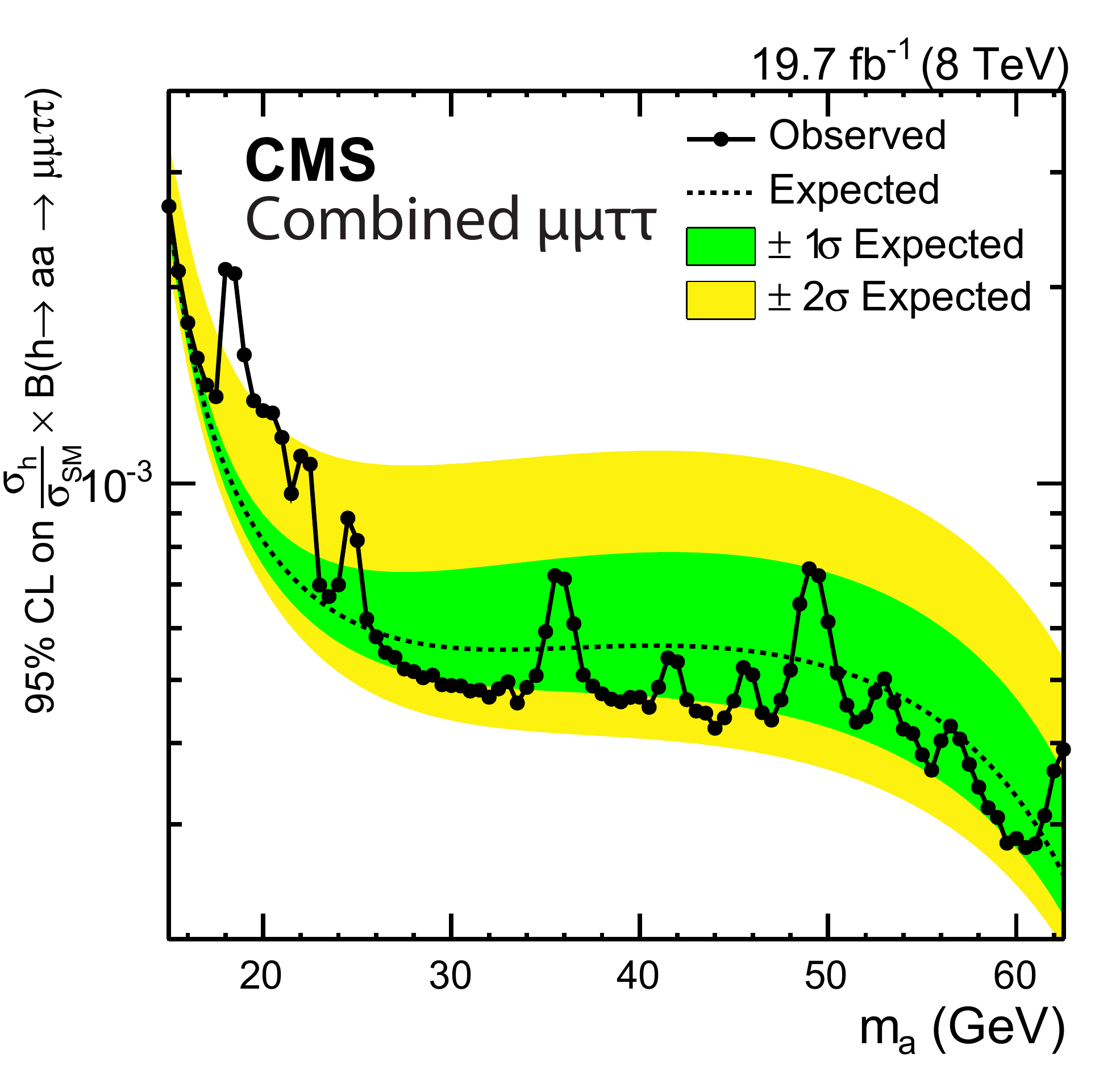}
\caption{Expected and observed upper limits at 95\% CL on the $\Hb$ boson production normalized to the SM prediction times $\mathcal{B}(\processmmtt)$
in the $\PGm^+\PGm^-\PGt_{\Pe}^+\PGt_{\Pe}^-$ (upper left), $\PGm^+\PGm^-\PGt_{\Pe}^\pm\PGt_{\PGm}^\mp$ (upper right), $\PGm^+\PGm^-\PGt_{\Pe}^\pm\tauh^\mp$ (middle left),
$\PGm^+\PGm^-\PGt_{\PGm}^\pm{\tauh}^\mp$ (middle right),
and $\PGm^+\PGm^-{\tauh}^+{\tauh}^-$ (lower left) final states, and for the combination of these five final states (lower right).
None of the event excesses exceed two standard deviations in global significance.
}
\label{fig:limits_mmtt}
\end{center}
\end{figure}

For the $\processmmtt$ analysis, upper limits on the production of $\Hb\to\ab\ab$ relative to
the SM Higgs boson production (including $\cPg\cPg\Hb$, VBF, $\PW\Hb$, $\PZ\Hb$, and $\cPqt\cPaqt\Hb$ production modes),
scaled by $\mathcal{B}(\ab\to\PGt^{+}\PGt^{-})\, \mathcal{B}(\ab\to\PGm^{+}\PGmm)$, are set.
An unbinned maximum likelihood fit to data is performed, and
upper limits are set at 95\% CL using the modified $\text{CL}_{\text{s}}$ method, taking into account the different yield and
shape systematic uncertainties described previously. The asymptotic approximation is not used in this case because of the low predicted background yields.
The limits are shown in Fig.~\ref{fig:limits_mmtt}. Considering the large look-elsewhere effect~\cite{Gross:2010qma} caused by the good dimuon mass resolution (about 2\%), the wide mass range probed, and the number of studied final states, none of the observed events corresponds to an excess of more than two standard deviations in global significance. In particular, the deviation of the observed limit with respect to the expected limit in the $\PGm^{+}\PGm^{-}\PGt_{\Pe}^\pm\PGt_{\PGm}^\mp$ final state comes from the presence of two observed events with a dimuon mass of 18.4 and 20.7\GeV, respectively, which lead to an excess of events with a maximum local significance of 3.5 standard deviations. Over the full mass range considered, the observed yield in this final state is compatible with the expected background yield of $1.80\pm0.74$ events. The uncertainty bands at low mass for most final states are narrow because of the low expected background yield.

\subsection{Interpretation of \texorpdfstring{$\Hb\to aa$}{haa} searches in 2HDM+S}

Searches for non standard decays of the SM-like Higgs boson to a pair of light pseudoscalar bosons are interpreted in the context of 2HDM+S.
In addition to the analyses presented in this paper, the results of two other searches are interpreted in this context:
the $\processmmmm$ search
covers pseudoscalar boson masses between 0.25 and 3.55\GeV~\cite{Khachatryan:2015wka},
whereas another $\processtttt$ search covers
pseudoscalar masses between 4 and 8\GeV with different boosted $\PGt$ lepton reconstruction techniques~\cite{CMStttt1}.
In 2HDM+S, the branching fractions of the light pseudoscalar $\ab$ to SM particles depend on the model type and on $\tan\beta$.
In type-1 2HDM+S, the fermionic couplings all have the same scaling with respect to the SM, whereas in type-2 2HDM+S (NMSSM-like),
they are suppressed for down-type fermions for
$\tan \beta<$ 1 (and enhanced for $\tan \beta>$ 1).
In type-3 2HDM+S (lepton specific), the decays to leptons are enhanced with respect to the decays to quarks for $\tan\beta>$ 1, and in type-4 2HDM+S (flipped),
the decays to up-type quarks and leptons are enhanced for $\tan\beta<$ 1.

\begin{figure}[htbp!]
\begin{center}
\includegraphics[angle=0,width=0.50\textwidth]{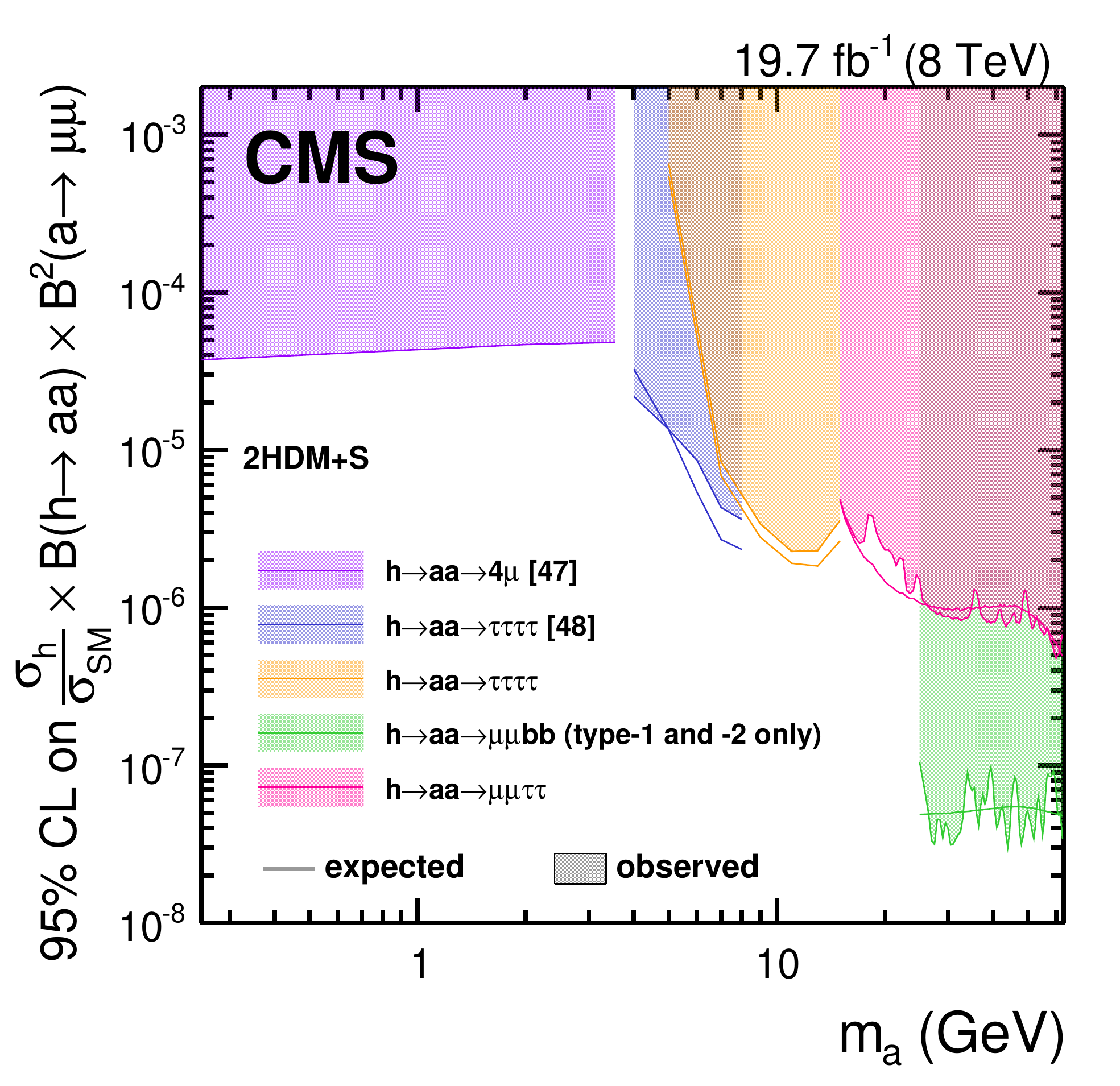}
\caption{Expected and observed 95\% CL exclusion limits on $({\sigma_{\Hb}}/{\sigma_{\textrm{SM}}}) \, \mathcal{B}(\Hb\to\ab\ab)\, \mathcal{B}^2(\ab\to\PGm^+\PGm^-)$ for various exotic $\Hb$ boson decay searches performed with data collected at 8\TeV with the CMS detector, assuming that the branching fractions of
the pseudoscalar boson to muons, $\PGt$ leptons and \PQb quarks follow Eqs.~(\ref{eq:2hdm})-(\ref{eq:2hdm1}). This assumption implies that the limit shown for $\processmmbb$ is valid only in type-1 and -2 2HDM+S.
}
\label{fig:compa_modelindependent}
\end{center}
\end{figure}

Because $\mathcal{B}(\ab\to\Pgt^{+}\Pgt^{-})$ is directly proportional to $\mathcal{B}(\ab\to\Pgm^+\Pgm^-)$ in any type of 2HDM+S as per Eq.~(\ref{eq:2hdm}), as is $\mathcal{B}(\ab\to \PQb\overline{\PQb})$ in type-1 and -2, the results of all analyses can be expressed as exclusion limits on $\frac{\sigma(\Hb)}{\sigma_{\textrm{SM}}} \, \mathcal{B}(\Hb\to \ab\ab)\, \mathcal{B}^2(\ab\to\Pgm^+\Pgm^-)$. This assumption is applied to obtain the results shown in Fig.~\ref{fig:compa_modelindependent}.
The exact value of $\mathcal{B}(\ab\to\PGm^+\PGm^-)$ depends on the type of 2HDM+S,
on $\tan\beta$ and on the pseudoscalar boson mass. No significant excess of events is observed for any of the five analyses. Under type-1 and -2 2HDM+S hypothesis, the $\processmmbb$ search is about one order of magnitude more sensitive than the $\processmmtt$ search, but does not cover the pseudoscalar mass range between 15 and 25\GeV. Both $\processtttt$ searches have a comparable sensitivity, in slightly different mass ranges.

\begin{figure}[htbp!]
\begin{center}
\includegraphics[angle=0,scale=0.38]{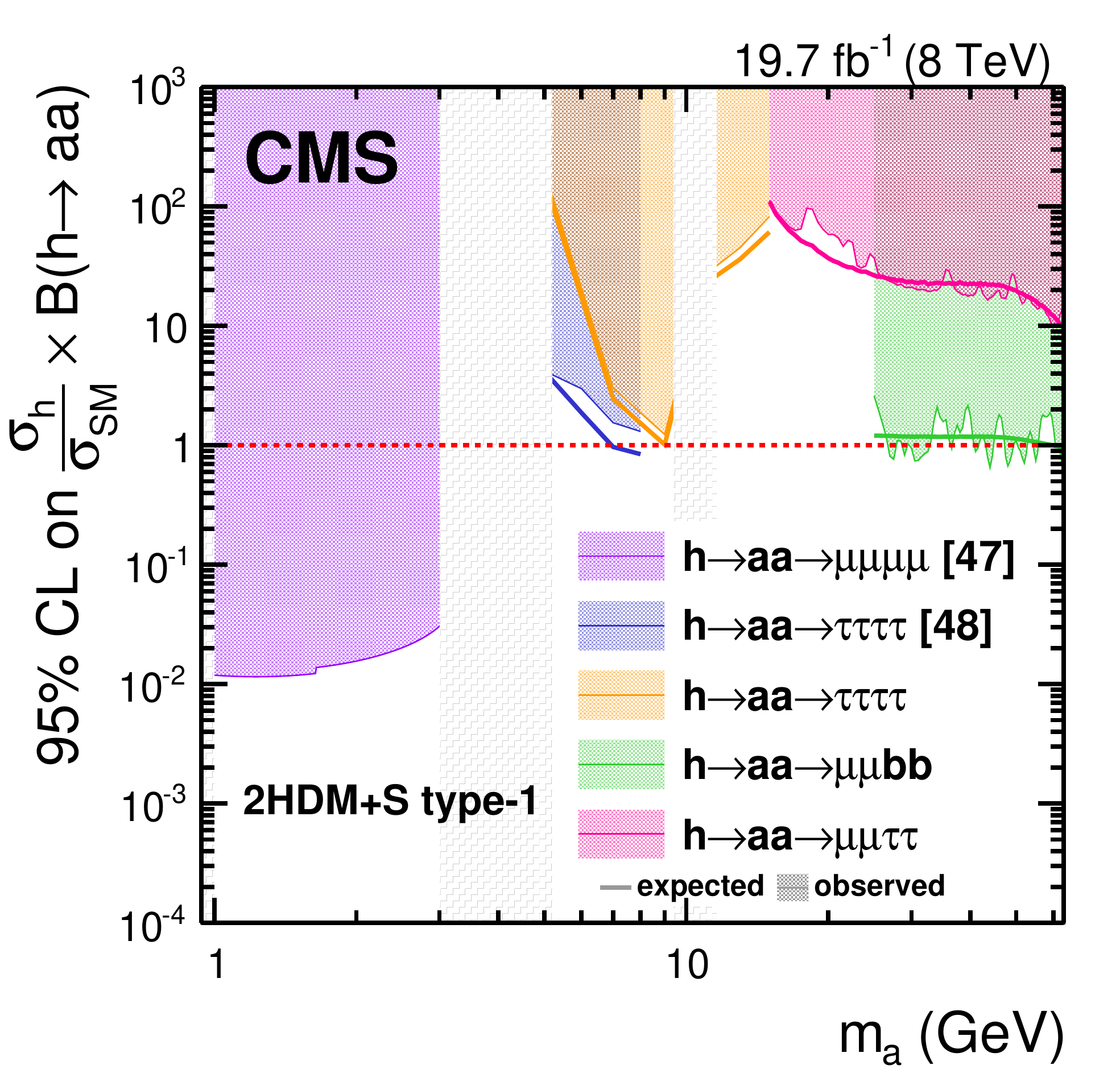}
\includegraphics[angle=0,scale=0.38]{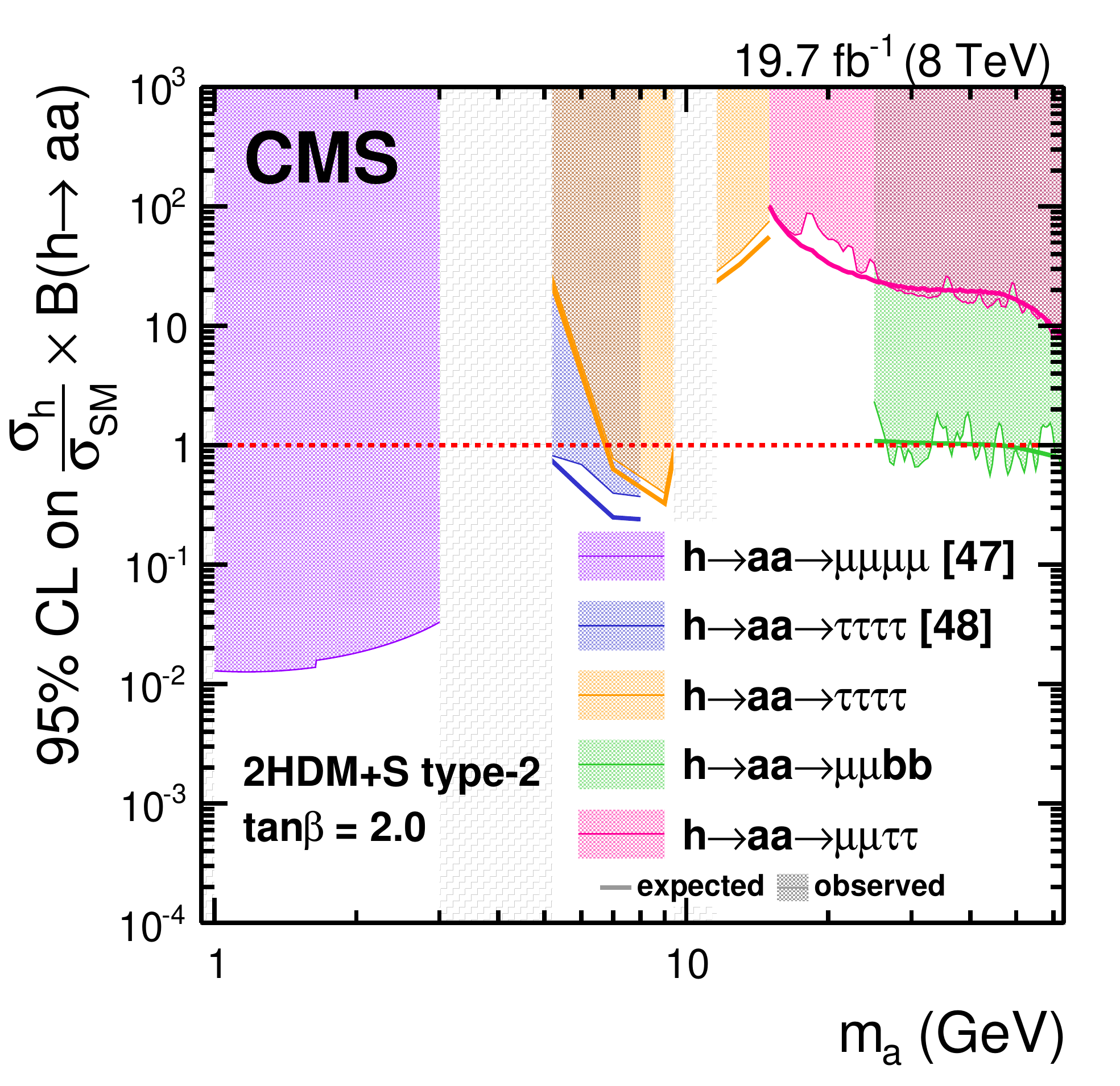}
\includegraphics[angle=0,scale=0.38]{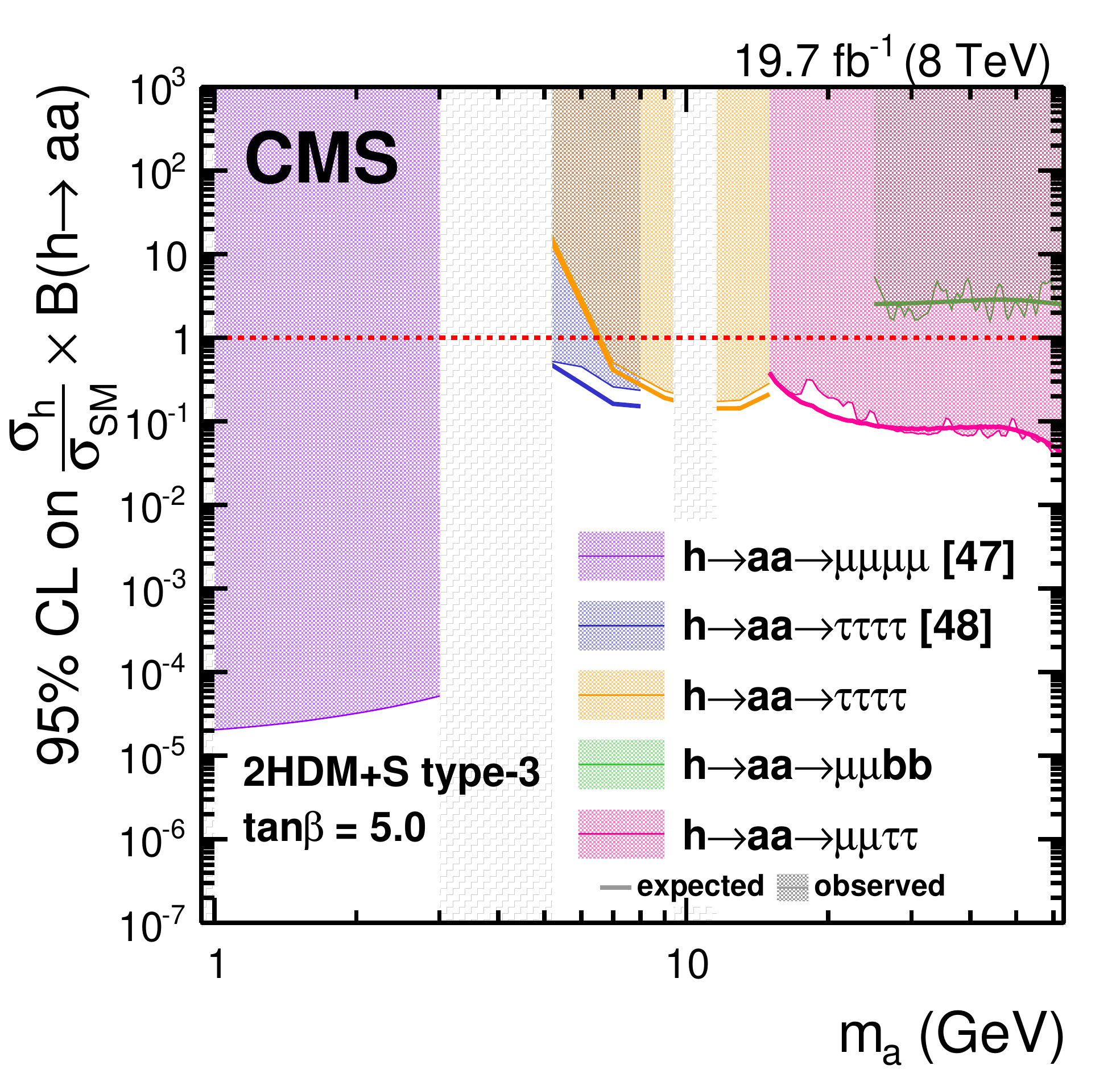}
\includegraphics[angle=0,scale=0.38]{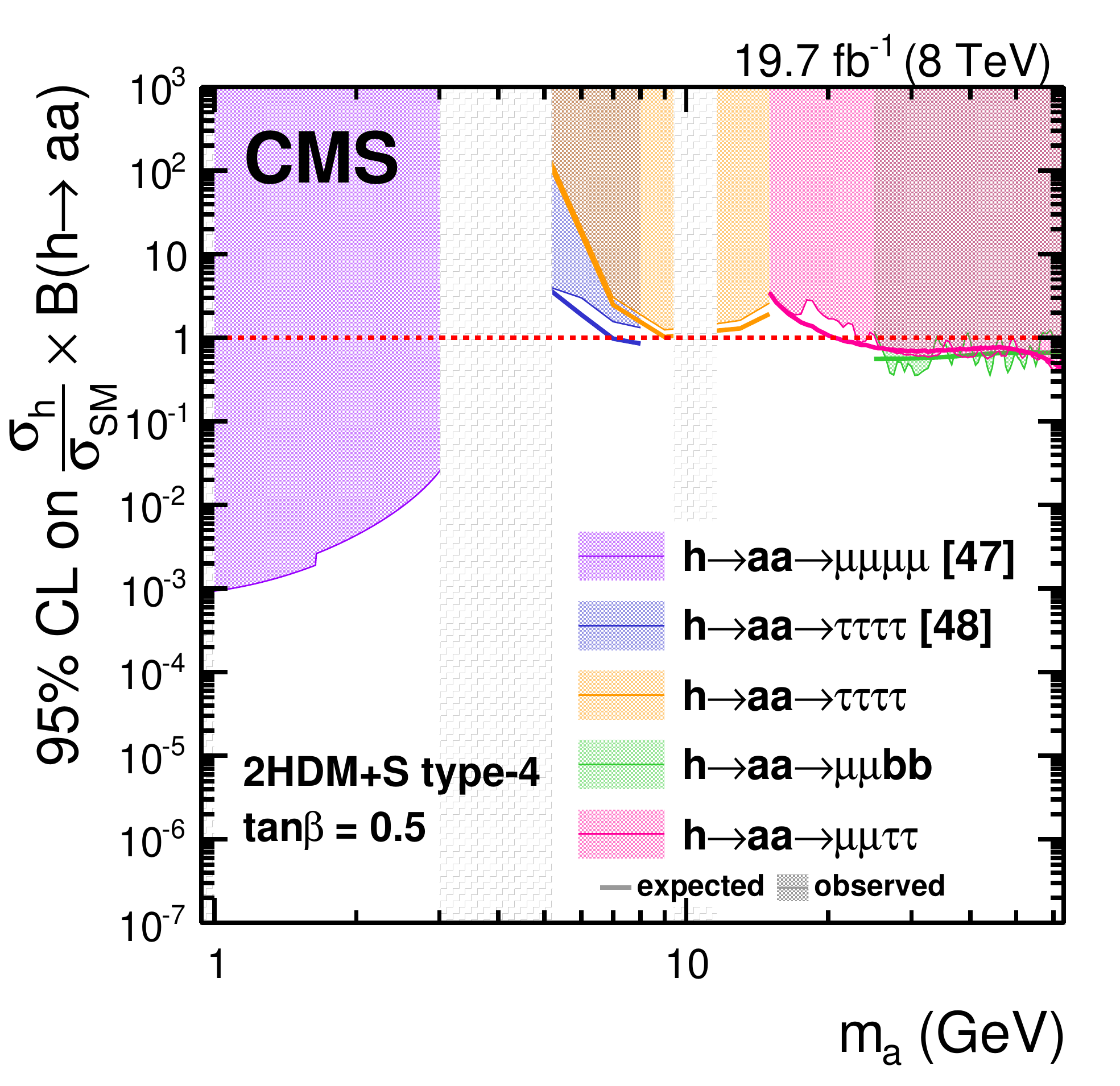}
\caption{Expected and observed 95\% CL limits on $( \sigma_{\Hb}/\sigma_{\textrm{SM}}) \, \mathcal{B}(\Hb\to\ab\ab)$ in 2HDM+S type-1 (top left),
type-2 with $\tan\beta=2$ (top right), type-3 with $\tan\beta=5$ (bottom left), and type-4 with $\tan\beta=0.5$ (bottom right). Limits are shown as a function of the mass of the light boson, $\ma$.
The branching fractions of the pseudoscalar boson to SM particles are computed
following a model described in Ref.~\cite{PhysRevD.90.075004}. Grey shaded regions correspond to regions where theoretical predictions for the branching fractions of the pseudoscalar boson to SM particles are not reliable.
}
\label{fig:compa_modeldependent}
\end{center}
\end{figure}

In 2HDM+S, the values of the branching fractions of the pseudoscalar boson to SM particles can be computed precisely, except for
pseudoscalar boson masses between approximately 3 and 5\GeV and  9 and 11\GeV because of decays to quarkonia, and for pseudoscalar boson masses
less than 1\GeV because of large QCD uncertainties in the hadronic final states~\cite{PhysRevD.90.075004}.
We compute them following the prescriptions in Refs.~\cite{PhysRevD.90.075004,anatomy}. The branching fractions used to interpret the results in the four particular 2HDM+S scenarios described below are given in Table~\ref{tab:BRaa}.
Figure~\ref{fig:compa_modeldependent} (top left) shows the 95\% CL in $({\sigma_{\Hb}}/{\sigma_{\textrm{SM}}}) \,\mathcal{B}(\Hb\to\ab\ab)$ in type-1 2HDM+S, for which there is no $\tan\beta$ dependence. Figure~\ref{fig:compa_modeldependent} (top right) shows corresponding limits in type-2 2HDM+S with $\tan\beta=2$; the sensitivity of the $\processtttt$ analyses is improved for $\ma<2\mb$ because of the enhancement of the couplings to leptons. The $\processtttt$ and $\processmmtt$ analyses have low sensitivity in type-1 2HDM+S and type-2 2HDM+S with $\tan\beta=2$ for $\ma>2\mb$, because, in these scenarios, decays to b quarks dominate over decays to $\PGt$ leptons and muons. The results in type-3 2HDM+S with $\tan\beta=5$ are depicted in the bottom left part of Fig.~\ref{fig:compa_modeldependent}; this scenario provides high sensitivity for the various analyses because of the enhancement of the couplings to leptons over those to quarks. Finally, the limits obtained in type-4 2HDM+S for $\tan\beta=0.5$ are shown in the bottom right part of Fig.~\ref{fig:compa_modeldependent}; the choice of $\tan\beta<1$ ensures large couplings to leptons. 
Regions where the theoretical predictions for the branching fractions of the pseudoscalar boson to SM particles are not reliable are indicated with grey shaded areas in the figure.
To obtain the exclusion limit for $\Hb\to\ab\ab\to 4\PGm$ in these hypotheses, the model-independent limit shown in Fig.~\ref{fig:compa_modelindependent}
is extrapolated from three mass points (0.25, 2.00, 3.55\GeV) to intermediate masses with a third degree polynomial, before being divided by the square of
$\mathcal{B}(\ab\to\PGm^+\PGm^-)$.
The variation of the limit around $\ma=1.5\GeV$, visible in Fig.~\ref{fig:compa_modeldependent}, is related to an increase of
the pseudoscalar boson decay width to gluons
because of the change in the number of active flavors in the QCD corrections and in the computation of the running of the strong
coupling constant at a renormalization scale equal to $\ma$.
The  $\PQb\overline{\PQb}\Hb$ production is neglected in this study. Its yield corresponds to less than 3\% of the total production cross section for $\tan\beta < 5$, but could be larger for higher $\tan\beta$ values, or due to other new physics effects.

\begin{table}[!htbp]
\renewcommand{\arraystretch}{1.2}
\begin{center}
\topcaption{Branching fractions of the pseudoscalar boson $\ab$ to muons, $\PGt$ leptons, and \PQb quarks, in the four 2HDM+S scenarios considered in Fig.~\ref{fig:compa_modeldependent}, as a function of the light boson mass. The branching fraction $\mathcal{B}(\ab\to\PQb\overline{\PQb})$ is not indicated in the mass range $\ma\in[5,15]\GeV$ because it is not used to interpret the results.}\label{tab:BRaa}
\resizebox{\textwidth}{!}{
\begin{tabular}{ll|c|c|c}
& & $\ma\in[1,3.5]\GeV$ & $\ma\in[5,15]\GeV$ & $\ma\in[20,62.5]\GeV$\\
\hline
\multirow{3}{*}{\begin{tabular}{l}Type-1\end{tabular}} & $\mathcal{B}(\ab\to\PGm^+\PGm^-)$ & $4.6\times 10^{-3}$ -- $4.0\times 10^{-2}$ & $2.1\times 10^{-4}$ -- $1.8\times 10^{-3}$ & $2.0\times 10^{-4}$ -- $2.2\times 10^{-4}$ \\
& $\mathcal{B}(\ab\to\PGt^+\PGt^-)$ & 0 & $5.7\times 10^{-2}$ -- $3.6\times 10^{-1}$ & $5.5\times 10^{-2}$ -- $6.3\times 10^{-2}$ \\
& $\mathcal{B}(\ab\to\PQb\overline{\PQb})$ & 0 & --- & $8.3\times 10^{-1}$ -- $8.8\times 10^{-1}$ \\
\hline
\multirow{3}{*}{\begin{tabular}{l}Type-2\\$\tan\beta=2$\end{tabular}}& $\mathcal{B}(\ab\to\PGm^+\PGm^-)$ & $2.5\times 10^{-2}$ -- $3.8\times 10^{-2}$ & $2.2\times 10^{-4}$ -- $4.0\times 10^{-3}$ & $2.1\times 10^{-4}$ -- $2.5\times 10^{-4}$ \\
 & $\mathcal{B}(\ab\to\PGt^+\PGt^-)$ & 0 & $6.0\times 10^{-2}$ -- $7.9\times 10^{-1}$ & $5.8\times 10^{-2}$ -- $7.0\times 10^{-2}$ \\
                       & $\mathcal{B}(\ab\to\PQb\overline{\PQb})$ & 0 & --- & $9.2\times 10^{-1}$ -- $9.3\times 10^{-1}$ \\
\hline
\multirow{3}{*}{\begin{tabular}{l}Type-3\\$\tan\beta=5$\end{tabular}}& $\mathcal{B}(\ab\to\PGm^+\PGm^-)$ & $7.4\times 10^{-1}$ -- $9.6\times 10^{-1}$ & $3.5\times 10^{-3}$ -- $5.0\times 10^{-3}$ & $3.4\times 10^{-3}$ -- $3.5\times 10^{-3}$ \\
& $\mathcal{B}(\ab\to\PGt^+\PGt^-)$ & 0 & $9.1\times 10^{-1}$ -- $9.9\times 10^{-1}$ & $9.7\times 10^{-1}$ \\
& $\mathcal{B}(\ab\to\PQb\overline{\PQb})$ & 0 & --- & $2.0\times 10^{-2}$ -- $2.5\times 10^{-2}$ \\
\hline
\multirow{3}{*}{\begin{tabular}{l}Type-4\\$\tan\beta=0.5$\end{tabular}} & $\mathcal{B}(\ab\to\PGm^+\PGm^-)$ & $4.5\times 10^{-3}$ -- $1.4\times 10^{-1}$ & $1.2\times 10^{-3}$ -- $1.8\times 10^{-3}$ & $1.1\times 10^{-3}$ -- $1.2\times 10^{-3}$ \\
& $\mathcal{B}(\ab\to\PGt^+\PGt^-)$ & 0 & $3.2\times 10^{-1}$ -- $3.5\times 10^{-1}$ & $3.0\times 10^{-1}$ -- $3.3\times 10^{-1}$ \\
& $\mathcal{B}(\ab\to\PQb\overline{\PQb})$ & 0 & --- & $2.5\times 10^{-1}$ -- $3.2\times 10^{-1}$ \\
\end{tabular}
}
\end{center}
\end{table}

\begin{figure}[htbp!]
\begin{center}
\includegraphics[angle=0,scale=0.38]{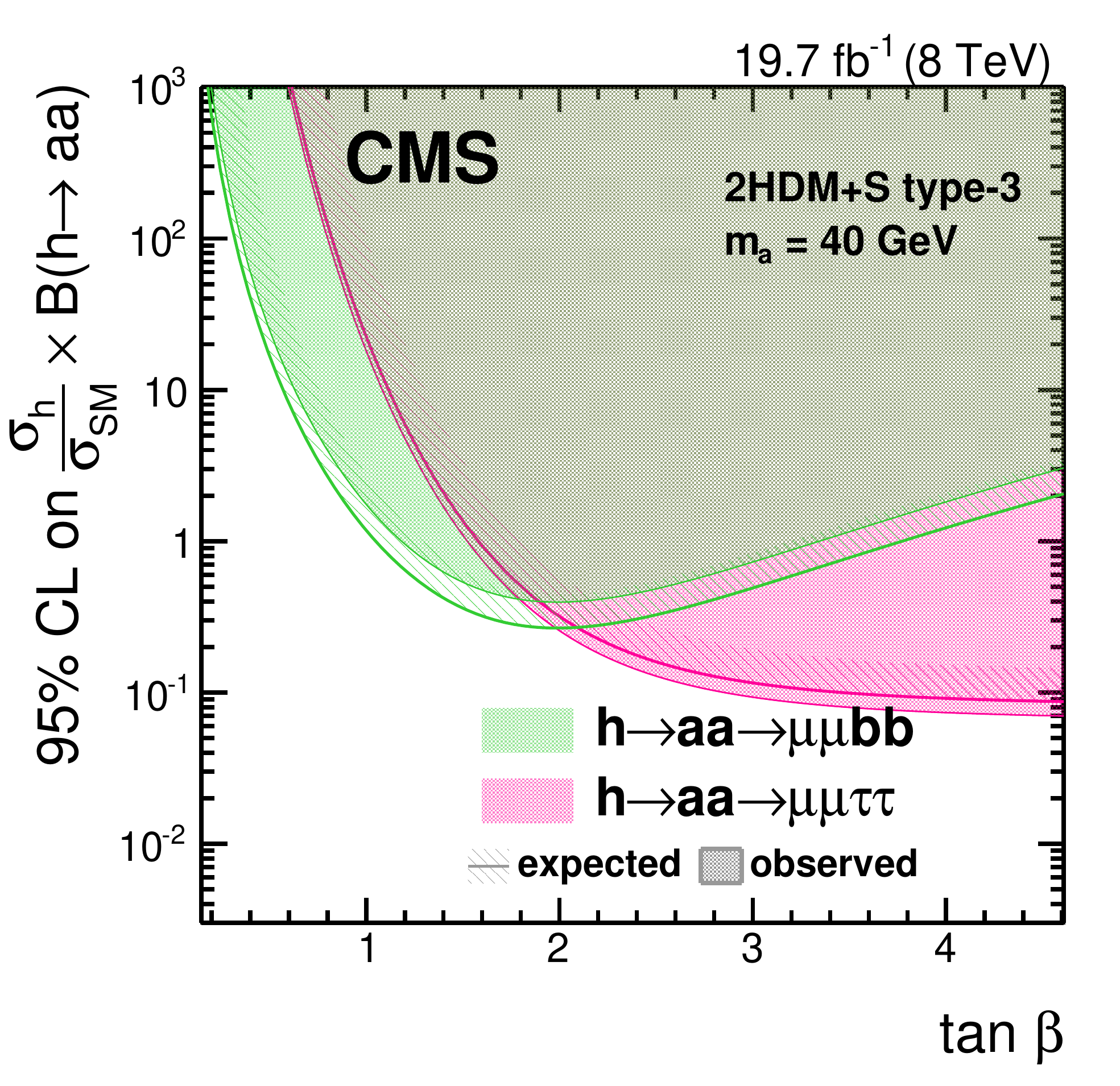}
\includegraphics[angle=0,scale=0.38]{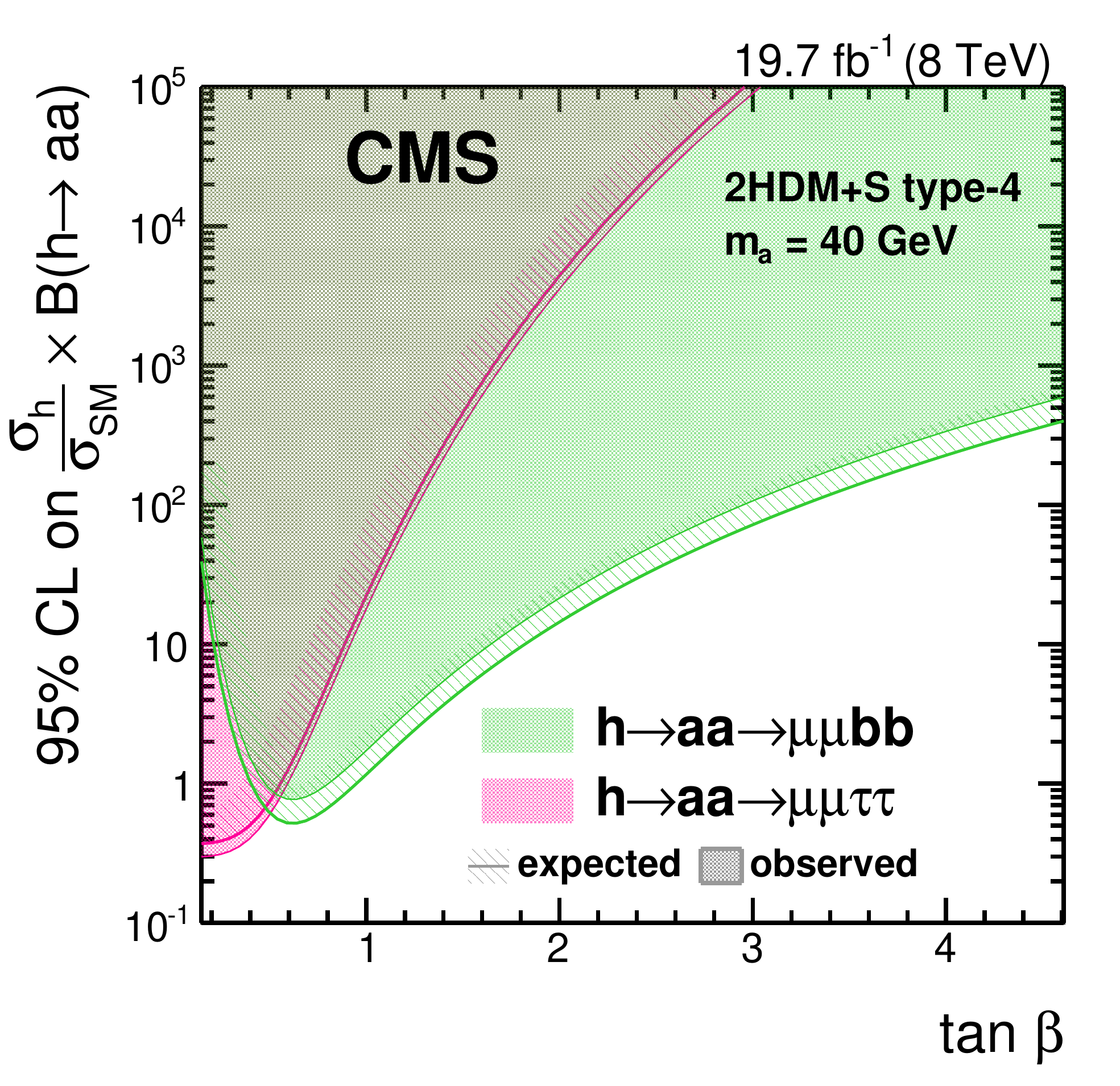}
\caption{The 95\% CL limit on $( \sigma_{\Hb}/\sigma_{\textrm{SM}} ) \, \mathcal{B}(\Hb\to\ab\ab)$ in 2HDM+S type-3 (left) and type-4 (right) for different
$\tan\beta$ values, for the $\processmmtt$ and $\processmmbb$ analyses at $\ma=40\GeV$.
The branching fractions of the pseudoscalar boson to SM particles are computed
following the prescriptions in Ref.~\cite{PhysRevD.90.075004}.
}
\label{fig:mmtt_mmbb}
\end{center}
\end{figure}

The $\processmmbb$ and $\processmmtt$ analyses are complementary over the $\tan\beta$ spectrum in
type-3 and -4 2HDM+S, where the ratio of the branching fractions of the pseudoscalar boson to $\PGt$ leptons and \PQb quarks depends on $\tan\beta$. The
former search is more sensitive in type-3 2HDM+S for $\tan\beta \lesssim 2.2$ and in type-4 2HDM+S for $\tan\beta\gtrsim 0.8$, as shown in Fig.~\ref{fig:mmtt_mmbb}.


The best limits on $\frac{\sigma_\Hb}{\sigma_{\textrm{SM}}} \, \mathcal{B}(\Hb\to \ab\ab)$ are obtained in type-3 2HDM+S with large $\tan\beta$ values for the $\processtttt$ and $\processmmtt$ analyses. As shown in Fig.~\ref{fig:compa_modeldependent} (bottom left), upper limits at 95\% CL as low as 17\% for the $\processtttt$ analysis and 4\% for the $\processmmtt$ analysis can be set for $\tan\beta=5$. Similarly low limits are achieved at higher $\tan\beta$.
The best limit for the $\processmmbb$ analysis is 16\%, and is obtained in type-3 2HDM+S too, but with $\tan\beta=2$ as shown in Fig.~\ref{fig:mmtt_mmbb} (left).  

\section{Summary}
Searches for the decay of the SM-like Higgs boson to pairs of light scalar particles have been performed using 19.7\fbinv of pp collisions at a
center-of-mass energy of 8\TeV,
collected by the CMS experiment at the LHC, in final states with $\PGt$ leptons, muons, or $\PQb$ quark jets.
Such signatures are motivated in light of the non-negligible branching fraction provided in recent experimental constraints for non-SM $\Hb$ decays.
The data were found to be compatible with SM predictions.
Whereas indirect measurements from the combination of data collected by the  ATLAS and CMS collaborations at the LHC at 8\TeV center-of-mass energy set an upper limit of 34\% on branching fraction of the Higgs boson to BSM, direct limits provide complementarity and improve the sensitivity to the 2HDM+S models for particular scenarios and pseudoscalar masses. Upper limits at 95\% CL on $({\sigma_{\Hb}}/{\sigma_{\textrm{SM}}}) \, \mathcal{B}(\Hb\to \ab\ab)$, assuming SM production of the 125\GeV Higgs boson, are as low as 17, 16, and 4\%, and have been determined for the $\processtttt$, $\processmmbb$, and $\processmmtt$ analyses, respectively.

\begin{acknowledgments}

\hyphenation{Bundes-ministerium Forschungs-gemeinschaft Forschungs-zentren Rachada-pisek} We congratulate our colleagues in the CERN accelerator departments for the excellent performance of the LHC and thank the technical and administrative staffs at CERN and at other CMS institutes for their contributions to the success of the CMS effort. In addition, we gratefully acknowledge the computing centers and personnel of the Worldwide LHC Computing Grid for delivering so effectively the computing infrastructure essential to our analyses. Finally, we acknowledge the enduring support for the construction and operation of the LHC and the CMS detector provided by the following funding agencies: the Austrian Federal Ministry of Science, Research and Economy and the Austrian Science Fund; the Belgian Fonds de la Recherche Scientifique, and Fonds voor Wetenschappelijk Onderzoek; the Brazilian Funding Agencies (CNPq, CAPES, FAPERJ, and FAPESP); the Bulgarian Ministry of Education and Science; CERN; the Chinese Academy of Sciences, Ministry of Science and Technology, and National Natural Science Foundation of China; the Colombian Funding Agency (COLCIENCIAS); the Croatian Ministry of Science, Education and Sport, and the Croatian Science Foundation; the Research Promotion Foundation, Cyprus; the Secretariat for Higher Education, Science, Technology and Innovation, Ecuador; the Ministry of Education and Research, Estonian Research Council via IUT23-4 and IUT23-6 and European Regional Development Fund, Estonia; the Academy of Finland, Finnish Ministry of Education and Culture, and Helsinki Institute of Physics; the Institut National de Physique Nucl\'eaire et de Physique des Particules~/~CNRS, and Commissariat \`a l'\'Energie Atomique et aux \'Energies Alternatives~/~CEA, France; the Bundesministerium f\"ur Bildung und Forschung, Deutsche Forschungsgemeinschaft, and Helmholtz-Gemeinschaft Deutscher Forschungszentren, Germany; the General Secretariat for Research and Technology, Greece; the National Scientific Research Foundation, and National Innovation Office, Hungary; the Department of Atomic Energy and the Department of Science and Technology, India; the Institute for Studies in Theoretical Physics and Mathematics, Iran; the Science Foundation, Ireland; the Istituto Nazionale di Fisica Nucleare, Italy; the Ministry of Science, ICT and Future Planning, and National Research Foundation (NRF), Republic of Korea; the Lithuanian Academy of Sciences; the Ministry of Education, and University of Malaya (Malaysia); the Mexican Funding Agencies (BUAP, CINVESTAV, CONACYT, LNS, SEP, and UASLP-FAI); the Ministry of Business, Innovation and Employment, New Zealand; the Pakistan Atomic Energy Commission; the Ministry of Science and Higher Education and the National Science Centre, Poland; the Funda\c{c}\~ao par\ab\ab Ci\^encia e a Tecnologia, Portugal; JINR, Dubna; the Ministry of Education and Science of the Russian Federation, the Federal Agency of Atomic Energy of the Russian Federation, Russian Academy of Sciences, the Russian Foundation for Basic Research and the Russian Competitiveness Program of NRNU MEPhI (M.H.U.); the Ministry of Education, Science and Technological Development of Serbia; the Secretar\'{\i}a de Estado de Investigaci\'on, Desarrollo e Innovaci\'on and Programa Consolider-Ingenio 2010, Spain; the Swiss Funding Agencies (ETH Board, ETH Zurich, PSI, SNF, UniZH, Canton Zurich, and SER); the Ministry of Science and Technology, Taipei; the Thailand Center of Excellence in Physics, the Institute for the Promotion of Teaching Science and Technology of Thailand, Special Task Force for Activating Research and the National Science and Technology Development Agency of Thailand; the Scientific and Technical Research Council of Turkey, and Turkish Atomic Energy Authority; the National Academy of Sciences of Ukraine, and State Fund for Fundamental Researches, Ukraine; the Science and Technology Facilities Council, UK; the US Department of Energy, and the US National Science Foundation.

Individuals have received support from the Marie-Curie program and the European Research Council and EPLANET (European Union); the Leventis Foundation; the A. P. Sloan Foundation; the Alexander von Humboldt Foundation; the Belgian Federal Science Policy Office; the Fonds pour la Formation \`a la Recherche dans l'Industrie et dans l'Agriculture (FRIA-Belgium); the Agentschap voor Innovatie door Wetenschap en Technologie (IWT-Belgium); the Ministry of Education, Youth and Sports (MEYS) of the Czech Republic; the Council of Science and Industrial Research, India; the HOMING PLUS program of the Foundation for Polish Science, cofinanced from European Union, Regional Development Fund, the Mobility Plus program of the Ministry of Science and Higher Education, the National Science Center (Poland), contracts Harmonia 2014/14/M/ST2/00428, Opus 2014/13/B/ST2/02543, 2014/15/B/ST2/03998, and 2015/19/B/ST2/02861, Sonata-bis 2012/07/E/ST2/01406; the Thalis and Aristeia programs cofinanced by EU-ESF and the Greek NSRF; the National Priorities Research Program by Qatar National Research Fund; the Programa Clar\'in-COFUND del Principado de Asturias; the Rachadapisek Sompot Fund for Postdoctoral Fellowship, Chulalongkorn University and the Chulalongkorn Academic into Its 2nd Century Project Advancement Project (Thailand); and the Welch Foundation, contract C-1845.

\end{acknowledgments}

\clearpage

\bibliography{auto_generated}
\cleardoublepage \appendix\section{The CMS Collaboration \label{app:collab}}\begin{sloppypar}\hyphenpenalty=5000\widowpenalty=500\clubpenalty=5000\textbf{Yerevan Physics Institute,  Yerevan,  Armenia}\\*[0pt]
V.~Khachatryan, A.M.~Sirunyan, A.~Tumasyan
\vskip\cmsinstskip
\textbf{Institut f\"{u}r Hochenergiephysik,  Wien,  Austria}\\*[0pt]
W.~Adam, E.~Asilar, T.~Bergauer, J.~Brandstetter, E.~Brondolin, M.~Dragicevic, J.~Er\"{o}, M.~Flechl, M.~Friedl, R.~Fr\"{u}hwirth\cmsAuthorMark{1}, V.M.~Ghete, C.~Hartl, N.~H\"{o}rmann, J.~Hrubec, M.~Jeitler\cmsAuthorMark{1}, A.~K\"{o}nig, I.~Kr\"{a}tschmer, D.~Liko, T.~Matsushita, I.~Mikulec, D.~Rabady, N.~Rad, B.~Rahbaran, H.~Rohringer, J.~Schieck\cmsAuthorMark{1}, J.~Strauss, W.~Waltenberger, C.-E.~Wulz\cmsAuthorMark{1}
% ...
\vskip\cmsinstskip
% ...
1:~~Also at Vienna University of Technology, Vienna, Austria\\
% ...

\end{sloppypar}
\end{document}